\documentclass[12pt,a4paper]{article}
\usepackage[centertags]{amsmath}
\usepackage{amssymb}
\usepackage{epsfig} 
\usepackage{amsmath} 
\usepackage{graphicx} 
\usepackage{jheppub}
\usepackage{slashed}

\newcommand{\beq}{\begin{equation}}
\newcommand{\eeq}{\end{equation}}
\newcommand{\bea}{\begin{eqnarray}}
\newcommand{\eea}{\end{eqnarray}}
\newcommand{\ba}{\begin{array}}
\newcommand{\ea}{\end{array}}
\newcommand{\bi}{\begin{itemize}}
\newcommand{\ei}{\end{itemize}}
\newcommand{\bn}{\begin{enumerate}}
\newcommand{\en}{\end{enumerate}}
\newcommand{\bc}{\begin{center}}
\newcommand{\ec}{\end{center}}
\renewcommand{\l}{\left}
\renewcommand{\r}{\right}

\newcommand{\Ga}{\Gamma}

\newcommand{\Om}{\Omega}

\newcommand{\al}{\alpha}

\newcommand{\la}{\lambda}

\newcommand{\si}{\sigma}

\newcommand{\nl}{\nonumber\\}

\newcommand{\eq}[1]{Eq.~(\ref{#1})}

\newcommand{\GeV}{\mathinner{\mathrm{GeV}}}
\newcommand{\TeV}{\mathinner{\mathrm{TeV}}}

\newcommand{\pl}{{\rm Pl}}

\def\gtsim{\mathrel{\hbox{\raise0.2ex
\hbox{$>$}\kern-0.75em\raise-0.9ex\hbox{$\sim$}}}}
\def\ltsim{\mathrel{\hbox{\raise0.2ex
\hbox{$<$}\kern-0.75em\raise-0.9ex\hbox{$\sim$}}}}
\newcommand{\Vac}[1]{\bigg\langle{#1}\bigg\rangle}

\title{Vacuum structure and stability of a singlet fermion dark matter 
model with a singlet scalar messenger}

\author[a]{Seungwon Baek,} 
\author[a]{P. Ko,}
\author[a]{Wan-Il Park}
\author[a]{and Eibun Senaha}

\affiliation[a]{School of Physics, KIAS, \\ Seoul 130-722, Korea}

\emailAdd{sbaek1560@gmail.com}
\emailAdd{pko@kias.re.kr}
\emailAdd{wipark@kias.re.kr}
\emailAdd{senaha@kias.re.kr}

\abstract{
We consider the issue of vacuum stability and triviality bound of
the singlet extension of the Standard Model (SM) with a singlet fermion 
dark matter (DM). In this model,  the singlet scalar plays the role of a 
messenger between the SM sector and the dark matter sector.
This model has two Higgs-like scalar bosons, and is consistent with 
all the data on electroweak precision tests, thermal relic density of 
DM and its direct detection constraints. 
We show that this model is stable without hitting Landau pole 
up to Planck scale for 125 GeV Higgs boson. 
We also perform a comprehensive study of vacuum structure, and point out 
that a region where electroweak vacuum is the global minimum is highly limited.
In this model, both Higgs-like scalar bosons have reduced couplings 
to the SM weak gauge bosons and the SM fermions,
because of the mixing between the SM Higgs boson and the singlet scalar. 
There is also a possibility of their invisible decay(s) into a pair of DM's.
Therefore  this model would be disfavored if the future data on the 
$( \sigma \cdot B )_{VV}$ or $( \sigma \cdot B )_{f\bar{f}}$ with 
$V=\gamma,W,Z$ and $f=b, \tau$ turn out larger than the SM 
predictions.}

\preprint{KIAS-P12061}


\begin{document}
\maketitle

\section{Introduction}
The Standard Model (SM) has been extremely successful in describing 
interactions between quarks and leptons down to $\sim 10^{-19}$ m,
or up to a few TeV depending on the channels,
and it would be complete if the Higgs sector is established experimentally,
clarifying the origin of electroweak symmetry breaking (EWSB) and masses 
for the SM chiral fermions and EW gauge bosons.  

However, the SM should be extended in order to accommodate 3 different 
directions:
\begin{itemize}
\item Neutrino masses and mixings cannot be explained within the context of 
renormalizable SM.
\item Baryon number asymmetry of the universe requires a new source of CP 
violation, beyond the CKM phase in the SM~\cite{SMewbg_CPV}.
\item Nonbaryonic cold dark matter of the universe should be included 
in the SM.
\end{itemize}

Admittedly, the simplest and the most economic solution to the 1st and the 2nd 
problems is to invoke the seesaw mechanism~\cite{seesaw} and leptogenesis by introducing 
extra singlet right-handed (RH) neutrinos~\cite{leptogenesis}. 
This is also nicely fit to the idea of grand unified theory (GUT) based on $SU(5)$ 
and $SO(10)$. 

For the nonbaryonic cold dark matter, there are many candidates in particle 
physics models: axion and axino, the lightest supersymmetric particle (LSP) 
in SUSY models (neutralino or gravitino), the lightest Kaluza-Klein particle  
(LKP) in extra dimensional scenarios, to name only a few. Some of them are 
related with other problems in particle physics, such as strong CP problem
or fine tuning problem of (Higgs mass)$^2$, but there are many other models
which are not related with other problems in particle physics. 

Another possibility is to rely on the principle of Occam's razor, namely the simplest 
extension of the SM with dark matter candidates.  In terms of the least number of 
new degrees of freedom, a scalar DM model with $Z_2$ symmetry would be 
the simplest one. However the origin of $Z_2$ symmetry is not clear, since it is 
put in by hand.  
The simplest DM model without ad hoc $Z_2$ symmetry would be a singlet
Dirac fermion CDM with conserved charge associated with a global dark 
$U(1)$ symmetry.
In Ref.~\cite{Baek:2011aa}, three of the present authors proposed such a scenario,
by considering a Dirac fermion DM ($\psi$) that couples to a real singlet scalar ($S$)
(see also \cite{Kim:2008pp} and \cite{LopezHonorez:2012kv} for similar discussions). 
Writing the most general renormalizable lagrangian among these new fields 
($\psi$ and $S$)  and the SM fields, including the so-called Higgs portal terms, 
we could describe the DM physics (thermal relic density and direct detection), 
and collider phenomenology and electroweak precision tests (EWPT).
Adding the singlet scalar improves the overall EWPT fits~\cite{Baek:2011aa}. 

In this model, the Higgs phenomenology is modified in an important way 
by two different reasons:
\begin{itemize}
\item There are two neutral Higgs-like scalar bosons, $H_1$ and $H_2$, 
which are two mixtures of the SM Higgs boson $h$ and a singlet scalar $s$, 
with a mixing angle $\alpha$.  Couplings of $H_1$ and $H_2$ to the SM 
particles are reduced by $\cos\alpha$ or $\sin\alpha$. 
Therefore the production cross sections for $H_{i=1,2}$'s at colliders will be 
reduced by $\cos^2 \alpha$ or $\sin^2 \alpha$ compared with that  
of the SM Higgs boson with the same mass.
\item Both $H_1$ and $H_2$ can decay invisibly into a pair of DM if kinematically 
allowed : $H_i \rightarrow \psi \overline{\psi}$. This would make more difficult to 
observe the $H_i$'s produced at colliders. 
\end{itemize}
These two independent mechanisms will make two Higgs-like scalar bosons 
have reduced signal strength $\sigma \cdot B$ into specific final states
[ see Eq. ~(3.3) ].  

Recently, ATLAS and CMS reported a tantalizing hint for a Higgs-like 
boson with mass around 125 GeV~\cite{higgsdiscovery,Higgs_ATLAS,Higgs_CMS}.  
Its couplings to the $WW$, $ZZ$ and $\gamma\gamma$ are consistent 
with the SM predictions,  albeit there are still large uncertainties because 
of limited statistics.  More data accumulation is planned toward the end of 
this year, and we would learn much more about the detailed properties 
of the observed new Higgs-like boson. 

If the SM Higgs boson has mass around 125 GeV, the electroweak (EW) 
vacuum might be meta-stable or even unstable due to the quantum corrections from top quark 
loop \cite{Krive:1976sg,Cabibbo:1979ay,Anderson:1990aa,EliasMiro:2011aa,Degrassi:2012ry}
though large uncertainties in determining SM quantities including top pole mass 
do not allow to draw  a firm conclusion on this issue \cite{EliasMiro:2011aa,Alekhin:2012py,Masina:2012tz}. 
Meta-stability might be still allowed as long as the tunneling time to wrong vacuum is longer than the age of our universe.
However, if the primordial inflation is supposed to take place along the direction of Higgs field, for example as the case of Higgs inflation \cite{Bezrukov:2007ep}, the possible meta-stability of EW vacuum should be improved \cite{Degrassi:2012ry}
\footnote{
Inflation due to Higgs field false vacuum \cite{Masina:2011un,Masina:2011aa,Masina:2012yd} might be a possible alternative to Higgs inflation though the initial condition for inflation looks non-trivial to be realized.
}.
Some new physics should be introduced well below the Planck scale
in order to save this situation.  It is the purpose of this paper to address
this issue within the model proposed in Ref.~\cite{Baek:2011aa}.  
In this model, there are only two more fields beyond the SM ones, 
the fermion DM $\psi$ and a real singlet scalar messenger $S$. 
Since $S$ couples to the SM Higgs field directly,  one can imagine that 
the EW vacuum in our model could be stable even if
the new physics scale $\Lambda$ is as large as Planck scale. In other words,
the Planck chimney could be possible for Higgs mass around 125 GeV.

However, once we introduce $S$ and $\psi$ into the SM, 
the vacuum structure can change significantly, which may give rise to various false vacua.
For example, if $S$ develops the vacuum expectation value (VEV), 
the Higgs potential in the $S$ direction could take the form of a double-well potential.
If it is tilted, it is no longer clear that the EW vacuum is the the global minimum.
Furthermore, at the loop level $\psi$ contributes to the Higgs potential and may affect the vacuum 
structure as well. Nevertheless, such a vacuum analysis was often overlooked in the literature.

In this paper, we investigate the vacuum structure and stability in the SM with $S$ and $\psi$.
An effective potential approach is adopted to study the vacuum structure.
We explore not only the EW vacuum but also other possible false vacua at the tree- and one-loop
levels. At the tree level, we explicitly derive analytic expressions for the vacuum energies  
while the one-loop analysis exclusively relies on numerics.

In order to examine the vacuum instability occurring at the high-energy scale, 
we use the renormalization group (RG) method.
The $\beta$ functions of all dimensionless couplings are derived at the one-loop level.
As for the most relevant parameters such as the top quark Yukawa coupling, strong coupling and
SU(2) doublet Higgs quartic coupling, we also include two-loop contributions coming from the SM sector.
In addition to the vacuum stability, we also investigate the perturbativity of the quartic couplings up to the Planck scale.

This paper is organized as follows. In Sec.~\ref{sec:model}, we describe the model and 
the relevant constraints from colliders and dark matter physics are discussed
in Sec.~\ref{sec:constraints}.  
In Sec.~\ref{sec:vac_struc}, we discuss the vacuum structures, and various vacua are scrutinized carefully. 
The stability of EW vacuum taking account of the RG effects up to some new physics scale $\Lambda$ 
is investigated in Sec.~\ref{sec:vac_stabil}.  
The paper is summarized in Sec.~\ref{sec:conclu}.  
We collect the matching conditions used in our analysis in Appendix~\ref{app:top_quark}, 
and the one-loop Higgs boson mass formulae as well as one-loop tadpole conditions are presented 
in Appendix \ref{app:1LmH}.

\section{The model}\label{sec:model}

We consider a SM gauge-singlet Dirac fermion DM ($\psi$) with a real singlet 
scalar ($S$) that couples to the SM sector by the Higgs portal \cite{Baek:2011aa}.
The dark sector is described by the lagrangian
\bea
 {\cal L}_{\rm dark} = \overline{\psi}(i \slashed \partial - m_{\psi_0}) \psi
 - \lambda S \overline{\psi} \psi \ .
\eea
The most general renormalizable scalar potential including the Higgs portal
interactions is given by
\bea
V &=& - \mu_H^2 H^\dag H + \lambda_H \l( H^\dag H \r)^2 
\\
&& + \mu_{HS} S H^\dag H + \frac{1}{2} \lambda_{HS} S^2 H^\dag H
\\
&& + \mu_S^3 S + \frac{1}{2} m_S^2 S^2 + \frac{1}{3} \mu_S' S^3 + 
\frac{1}{4} \lambda_S S^4,
\eea 
where $H$ is the SM Higgs field.

In general, the neutral scalar fields develop nontrivial 
vacuum expectation values (VEVs), $v_H$ and $v_S$. 
And we expand the neutral component of $H$ and $S$ as
\bea
H= \l(\begin{array}{c}  0 \\ {1 \over \sqrt{2} }  (v_H + h)  \end{array} \r), 
 \quad
 S = v_S + s,
\eea
in the unitary gauge.
Then the minimization conditions of the 
Higgs potential at VEVs give
\bea
 \mu_H^2 &=& \la_H v_H^2 + \mu_{HS} v_S 
 + {1 \over 2}  \la_{HS} v_S^2, \nl
 m_S^2 &=& -\frac{\mu_S^3}{v_S} - \mu_S^\prime v_S 
 - \la_S v_S^2 -\frac{\mu_{HS} v_H^2}{2 v_S}
 -{1 \over 2} \la_{HS} v_H^2.
\label{tree-tad}
\eea
We introduce the Higgs mixing angle $\alpha$ and the mass eigenvalues $m_{i(=1,2)}$
($m_1 < m_2$) which diagonalize the Higgs mass squared matrix such that
\bea
 M^2_{\rm Higgs} \equiv
 \l(\begin{array}{cc} m_{hh}^2 & m_{hs}^2 \\ m_{hs}^2 & m_{ss}^2 \end{array}\r)
 \equiv \l(\begin{array}{cc} \cos\al & \sin\al \\ -\sin\al & \cos\al \end{array}\r)
\l(\begin{array}{cc} m_1^2 & 0 \\ 0 & m_2^2 \end{array}\r)
\l(\begin{array}{cc} \cos\al & -\sin\al \\ \sin\al & \cos\al \end{array}\r).\label{conversion}
\eea
The quartic couplings, $\lambda_H, \lambda_{HS}, \lambda_{SS}$, in the Higgs potential
can be expressed in terms of the Higgs mass parameters
\bea
 \la_H &=& \frac{m_{hh}^2}{2 v_H^2}, \nl
 \la_{HS} &=& \frac{m_{hs}^2 -\mu_{HS} v_H}{v_S v_H}, \nl
 \la_{S} &=& \frac{m_{ss}^2 + \mu_S^3/v_S  + \mu_{HS} v_H^2/(2 v_S) 
 - \mu_S^\prime v_S}{2 v_S^2},\label{lam_trade}
\eea
so that they are obtained as a function of $m_1, m_2$ and $\alpha$ which
we take as input parameters.
The mass eigenstates $H_i$ ($i=1,2$) with masses $m_i$ are written
in terms of the SM Higgs scalar $h$ and the singlet scalar $s$  as
\bea
 H_1 &=& h \cos\al - s \sin\al, \nl
 H_2 &=& h \sin\al + s \cos\al.
\label{eq:mass_weak}
\eea
In the Higgs and dark sector we have 10 free parameters which are to 
be measured in the  experiments: 
\bea
 m_1, \quad m_2, \quad \al, \quad v_H, \quad v_S, \quad \mu_S, \quad 
\mu_S^\prime,\quad  \mu_{HS}, 
\quad  m_{\psi} (\equiv m_{\psi_0} + \la v_S), \quad \la.
\eea
The constraints on these parameters from perturbative unitarity of 
electroweak gauge boson scattering amplitudes, EWPT, collider searches for 
Higgs boson(s), DM relic density, DM direct detection experiments, 
were given in Ref.~\cite{Baek:2011aa}. And those from the vacuum stability
and triviality will be considered in this paper.

\section{The constraints considered in Ref.~\cite{Baek:2011aa}}\label{sec:constraints}

In Ref.~\cite{Baek:2011aa} we considered the following observables which
can constrain our model:
\begin{itemize}
\item the perturbative unitarity condition on the Higgs sector~\cite{bwlee,Englert:2011yb},
\item the LEP bound on the SM Higgs boson mass~\cite{Barate:2003sz},
\item the oblique parameters $S$, $T$ and $U$ obtained from the EWPT~\cite{Peskin:1990zt,Maksymyk:1993zm},
\item the observed CDM relic density, 
$\Omega_{\rm CDM} h^2 =0.1123 \pm 0.0035$~\cite{Jarosik:2010iu}, 
which we assume is saturated by the thermal relic $\psi$,
\item the upper bound on the DM-proton scattering cross section 
obtained by the XENON100 experiment~\cite{Aprile:2012nq}.
\end{itemize}

The first three conditions do not constrain the dark matter sector, 
and they are also relevant to the singlet scalar extension of the SM 
without dark matter.

\subsection{Perturbative unitarity of gauge boson scattering amplitudes}

The perturbative unitarity of  scattering amplitudes for longitudinal weak 
gauge bosons in our model  requires~\cite{bwlee,Englert:2011yb}, 
\beq 
\langle m^2 \rangle \equiv m_1^2 \cos^2 \alpha  + m_2^2 \sin^2 \alpha  \leq 
\frac{4 \pi \sqrt{2}}{3 G_F} \approx \left( 700 \GeV \right)^2 ,
\label{eq:unitarity}
\eeq 
If $m_1 \neq m_2$, \eq{eq:unitarity} can be re-expressed as
\beq
\sin \alpha^2 \leq \l( \frac{4 \pi \sqrt{2}}{3 G_F} - m_1^2 \r) / \l( m_2^2 - m_1^2 \r)
\eeq
which provides an upper-bound of the mixing angle as a function of $m_2$ 
for a given $m_1(\approx 125 {\rm GeV})$.

\subsection{Collider bound}

In Ref.~\cite{Baek:2011aa} we defined two ratios $r_i$ ($i=1,2$) 
(what we called the reduction factor):
\beq
r_i \equiv \frac{\si_{H_i} B_{H_i \to X_{\rm SM}}}{\si^{\rm SM}_{H_i} 
B^{\rm SM}_{H_i \to X_{\rm SM}}}\ \ (i=1,2),
\eeq
where $X_{\rm SM}$ is a specific SM final state, which measure the reduced
signal strength with respect to the SM.
In terms of the {\it SM Higgs (singlet Higgs)} decay width $\Ga_{H_i}^{\rm SM (hid)}$ 
with mass $m_i$, {\it i.e.} without the effect of the mixing, we get
\bea
 r_1 &=&  \frac{c_\al^4 \Ga_{H_1}^{\rm SM}}{c_\al^2 \Ga_{H_1}^{\rm SM}+
 s_\al^2 \Ga_{H_1}^{\rm hid}}, \nl
 r_2 &=&  \frac{s_\al^4 \Ga_{H_2}^{\rm SM}}{s_\al^2 \Ga_{H_2}^{\rm SM}+
 c_\al^2 \Ga_{H_2}^{\rm hid}+\Ga_{H_2\to H_1 H_1}}.
\label{eq:rf}
\eea
We can see that the ``reduction'' of signal strength is a generic feature
of this model, {\it i.e.} $r_i <1$. If the future LHC data on $r_1$ for some
$X_{\rm SM} = VV, f\overline{f}$ ($V=\gamma, W, Z$, $f=b, \tau$) is larger than
1, our model would be ruled out.  In Ref.~\cite{Baek:2011aa}, 
we found that if $r_1 \gtrsim 0.7$, we get $r_2 \lesssim 0.2$
for ($m_2 > m_1 (\approx 125 {\rm GeV})$). So the heavy
scalar boson will easily evade the detection at LHC.

If the Higgs splitting mode $H_2 \rightarrow H_1 H_1$  opens kinematically, it
would provide a smoking gun signal for our model, like the four $b$-jets,
two-photons plus two $b$-jets, four tau leptons, {\it etc}. The high-luminosity LHC
machine can target these signals.

\subsection{The oblique parameters: $S, T, U$}

\begin{figure}
\centering
\includegraphics[width=0.7\textwidth]{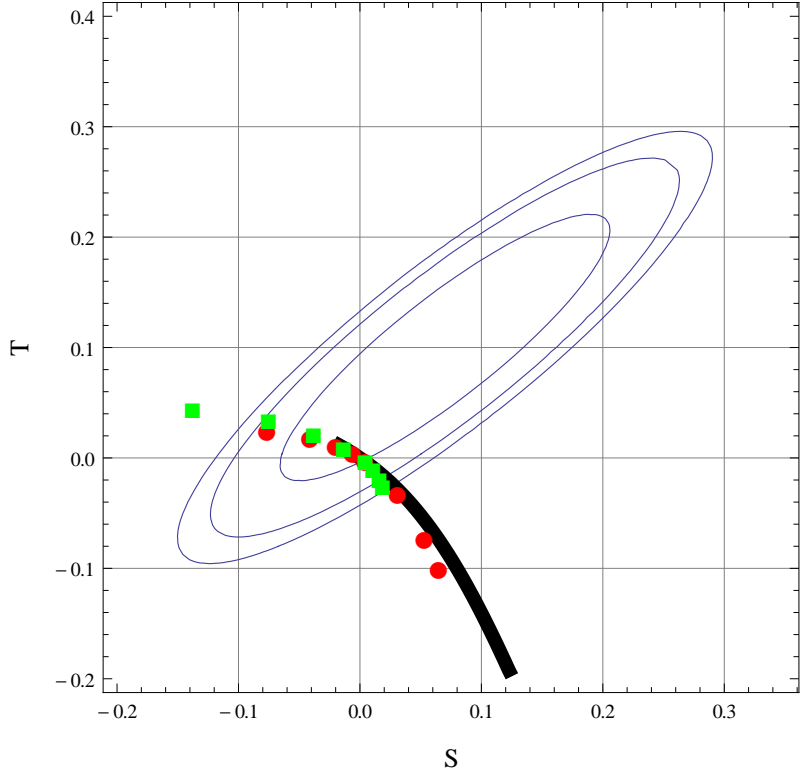}
\caption{
The prediction of $(S,T)$ parameters. We fixed the reference Higgs mass
to be 120 GeV.
The ellipses are (68, 90, 95)~\% CL contours from the global fit.
The thick black curve shows the SM prediction with the Higgs boson mass 
in the region $(100,720)$ GeV.
The red, green dots correspond to $\alpha=45^\circ, 20^\circ$, respectively.
The dots are for the choices $(m_1,m_2)(\GeV) =(25,125),(50,125),(75,125),(100,125),
(125,125),(125,250),(125,500),(125,750)$ from above for each color.
}
\label{fig:ST125}
\end{figure}

In our model the new scalar particle $S$ can contribute to the $W$ and $Z$ boson 
self-energy diagrams, $\Pi_{WW}, \Pi_{ZZ}$, thereby changing 
the EWPT $S, T, U$ parameters~\cite{Baek:2011aa}.
Explicit expressions for the oblique parameters in our model can be found in 
Ref.~\cite{Baek:2011aa}. The result is that including singlet scalar improves 
the overall fit to the EWPT, which can be seen 
in Fig.~\ref{fig:ST125}. 
For $m_1 \approx 125$ GeV, the mixing angle is constrained to be 
$\alpha \lesssim 0.4$ when $m_2 \gtrsim 400$ GeV.

\subsection{Dark matter relic density}

The observed DM relic density,
$\Omega_{\rm CDM} h^2 \simeq 0.1123 \pm 0.0035$~\cite{Jarosik:2010iu},
is related with the thermally averaged annihilation cross section times relative
velocity at freeze-out temperature roughly by\footnote{There is a typo in 
this expression in Ref.~\cite{Baek:2011aa} and we correct it here} 
\beq 
\Omega_{\rm CDM} h^2 \approx \frac{3 \times 10^{-27} {\rm cm^3/s}}
{\langle \sigma_{\rm ann} v \rangle_{\rm fz}}.
\label{thermal-average-sv}
\eeq

\begin{figure}
\centering
\includegraphics[width=\textwidth]{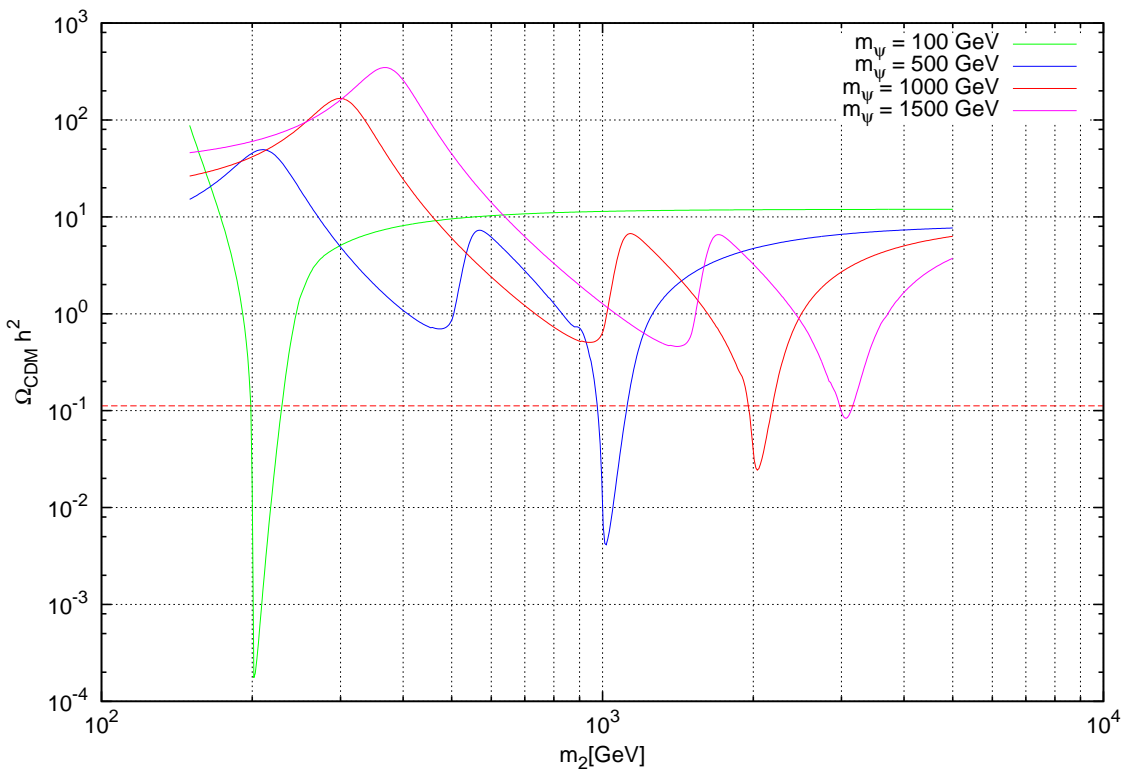}
\caption{Dark matter thermal relic density ($\Omega_{\rm CDM} h^2$) 
as a function of $m_2$ for $m_1=125 \GeV$, $\lambda=0.4$, $\alpha=0.1$ 
and $m_\psi = 100, 500, 1000, 1500 \GeV$ from top to bottom at right side.
The dotted red line corresponds to the observed value, $\Omega_{\rm CDM} h^2 = 0.112$.}
\label{fig:relic-density}
\end{figure}

The annihilation cross section of a DM pair is proportional to 
$\sin^2 2 \alpha$. Since the EWPT and LHC observation of the SM-like Higgs boson
restricts $\alpha$ to be small, the cross section is generically much smaller
than is needed to explained the current relic density. This can be seen in
Fig.~\ref{fig:relic-density} except for resonance regions.

At resonance, if $\gamma_i \equiv m_i \Gamma_i / ( 4 m_\psi^2 ) \ll 1$, 
the non-relativistic approximation of the cross-section is 
\cite{Gondolo:1990dk}
\bea 
\langle \sigma_{\rm res} v_{\rm lab} \rangle_{\rm NR} 
&=& \frac{4 \pi}{m_\psi^2} x^{3/2} \pi^{1/2} \gamma_i \, e^{-x \epsilon_i} 
\frac{B_i \left( 1 - B_i \right) \left( 1 + \epsilon_i \right)^{1/2}}{\left( 1 + 2 \epsilon_i \right)} \theta(\epsilon_i)
\eea
where $x \equiv m_\psi / T_{\rm fz}$ and 
$\epsilon_i \equiv -1 + m_i^2 / (4 m_\psi^2)$. 
For example, ignoring $H_2 \to H_1 H_1$ decay, we find 
\beq
\Gamma_i B_i \l( 1 - B_i \r) = \frac{\Gamma_i^{\rm SM} \Gamma_i^{\rm hid}}{\Gamma_i^{\rm SM} + \Gamma_i^{\rm hid}} ,
\eeq
If $m_2$ decays dominantly to dark matter,
\bea
\Gamma_{H_2} B_{H_2} \l( 1 - B_{H_2} \r) 
&\simeq& \frac{3 \sqrt{2}}{8 \pi} \sin^2 \alpha \ G_F m_f^2 m_2
\\
&=& \frac{3}{8 \pi} \sin^2 \alpha \  \l( \frac{m_t}{v} \r)^2 m_2 ,
\eea
and hence
\bea \label{s-sigmav-res}
\langle \sigma_{\rm res} v_{\rm lab} \rangle_{\rm NR} 
&\sim& \frac{3}{2} \pi^{1/2} x^{3/2} e^{-x \epsilon_{H_2}} 
\sin^2 \alpha \l( \frac{m_2}{ 2 m_\psi} \r)^2 \l( \frac{m_t}{v} \r)^2 
\frac{1}{m_\psi^2}
\\
&\simeq& 6 \times 10^{-5}  \sin^2 \alpha \l( \frac{1 \TeV}{m_\psi} \r)^2~{\rm GeV^{-2}}
\eea
where we have used $x = 25$ and $\epsilon_i = 1/x$ in the second line.	
Hence, even if $\alpha$ might be constrained to be small, a right amount 
of CDM relic density can be obtained as long as $m_\psi$ is in the band of 
$s$-channel resonance.   

We used the micrOMEGAs package~\cite{Belanger:2008sj} for numerical 
calculation of DM relic density and direct detection cross section. 
In Fig. \ref{fig:relic-density}, we show the CDM relic 
density as  a function of $m_2$ for various choices of $m_\psi = 100, 500, 
1000, 1500$ GeV, with $\lambda = 0.4$ and $\alpha = 0.1$. 
We can always find out the $m_2$ value which can accommodate thermal  
relic density of the singlet fermion CDM $\psi$. Note that there is no strong
constraint on the heavier Higgs with a small mixing angle $\alpha$, because 
$H_2$ would be mostly a singlet scalar so that it is very difficult to produce 
it at colliders, and also it could decay into a pair of CDM's with a substantial
branching ratio.

\subsection{Direct detection}

The null searches of the dark matter-proton scattering puts strong bounds 
on its spin-independent (SI) cross section~\cite{Aprile:2012nq}:
\beq 
\sigma_p \lesssim 2-10  \ (10^{-9} {\rm pb} )
\label{eq:XENON100}
\eeq
for the CDM in the mass range $m_{\rm CDM} = \mathcal{O}(10-100) \GeV$.
The spin-dependent scattering cross section is zero in our model, because the 
scattering is due to two Higgs-like scalar bosons.
Since the $\sigma_p$ is proportional to the $\langle
\sigma_{\rm ann} v \rangle_{\rm fz}$, the large annihilation cross section
need for the relic density would also give large DM-proton scattering
cross section, violating (\ref{eq:XENON100}).

As pointed out in \cite{Baek:2011aa},  there would be 
a destructive interference  between $H_1$ and $H_2$ contributions to 
the scattering amplitude due to orthogonality of the Higgs mixing matrix, 
which is a very generic aspect in case there are extra singlet scalar bosons that 
can mix with the SM Higgs boson \cite{progress}.
Hence, for regions $m_2 - m_1 \ll m_1$, a cancellation occurs in $\si_p$ 
and even the large $\la$, $\al$ regions are only weakly constrained 
(see Fig.~4 of Ref.~\cite{Baek:2011aa}).  
We also note that  $\sigma_p$ and $\langle \sigma_{\rm ann} v \rangle_{\rm fz}$ 
are not strongly correlated near the Higgs resonance where the relic density 
can be explained.  This helps to evade the strong  bound on $\sigma_p$ 
while accommodating the correct CDM density in the universe.   
This opens up a very interesting parameter space for Higgs boson search at the LHC,  
making one or two of the Higgs-like scalar  bosons can decay into a pair of DM's 
with a substantial invisible branching ratio(s).   

\subsection{Comparison with the effective lagrangian approach}
In this subsection, we would like to compare our model with the so-called 
Higgs portal fermion dark matter model~\cite{Djouadi:2011aa,LopezHonorez:2012kv},  where the singlet 
scalar $S$ is presumed to be integrated out, resulting in the following model 
lagrangian:
\begin{equation}
{\cal L}_{\rm eff} = \overline{\psi} \left( m_0 + \frac{H^\dagger H}{\Lambda} \right) \psi .
\end{equation}
Within this model, there is only one Higgs boson and its coupling to the 
DM is strongly constrained by the direct detection experiments. 
This result is very different from our analysis \cite{Baek:2011aa}, 
where there is a generic cancellation between $H_1$ and $H_2$ contributions 
in the direct detection rates.  In fact,  $\sigma_{\rm SI}$ depends also on 
$ ( \sin\alpha \cos\alpha )^2$, and it becomes zero when we ignore the mixing
between the SM Higgs boson and the singlet scalar $S$ 
(see Eq. (3.16) of Ref.~\cite{Baek:2011aa}).  This result can never be obtained in the 
approach based on the above effective lagrangian (3.13). 
In our case the correlation between $H_i - \psi - \overline{\psi}$ coupling and the direct detection 
cross section is not that strong compared with the results in Ref.~\cite{Djouadi:2011aa}.
It is important to consider the renormalizable models in order to discuss 
phenomenology related with the singlet fermion dark matter and Higgs bosons. 

The same arguments also applies to the Higgs portal vector  DM models, which is
assumed to be described by the following lagrangian:
\begin{equation}
{\cal L} = - m_V^2 V_\mu V^\mu - \frac{\lambda_{VH}}{4} H^\dagger H V_\mu V^\mu 
- \frac{\lambda_V}{4} ( V_\mu V^\mu )^2 \ .
\end{equation}
Although this lagrangian looks power-counting renormalizable, it is not really 
renormalizable. This is well known from the old intermediate vector boson theory
for weak gauge boson $W^\pm$.  In order to give a mass to a spin-1 gauge boson,
we need some symmetry breaking agency. Assuming a new complex scalar $\phi_X$
breaks the gauge symmetry spontaneously, one ends up with a new scalar boson 
from $\phi_X$ which would mix with the SM Higgs boson by Higgs portal. 
Therefore there will be two Higgs-like scalar boson in the end, and phenomenology in
the scalar sector should be similar to that of the model described here and in 
Ref.~\cite{Baek:2011aa}.   We leave the detailed discussions of this issue for the future 
publication~\cite{progress}. 

\section{Vacuum structure}\label{sec:vac_struc}
Because of the presence of the singlet scalar, the vacuum structure of this model 
is not that trivial.  
Since the Higgs potential is the quartic function of the Higgs fields (at the tree level), 
there could be another nondegenerate local minimum in the singlet Higgs direction unless some symmetry exists.
If that is the case, our EW vacuum may not be global and its stability is unclear.
In addition to this, the EW vacuum could be destabilized at a high energy scale 
by the RG effect of the top quark as in the SM. 
We separately examine the vacuum stability at the EW scale and the high energy scale.
In this section, we focus on the former, and the latter will be discussed in the next section.
%
%
\subsection{Tree level analysis}
Let us first consider the vacuum structures of our model at tree level
\footnote{
The vacuum analysis of the 
singlet extension of the SM within the electroweak phase transition context 
can be found in \cite{Espinosa:2011ax}.
}.
In this analysis, $\mu_H^2$ and $m_S^2$ are determined by Eq.~(\ref{tree-tad}) 
with fixed $v_H$ and $v_S$. 
The tree-level effective potential then takes the form
\begin{align}
V_0(\varphi_H, \varphi_S) 
&= \frac{\lambda_H}{4}(\varphi_H^4-2v_H^2\varphi_H^2)
	+\frac{\mu_{HS}}{2}
	\left(\varphi_H^2\varphi_S-\varphi_H^2v_S-\frac{1}{2}
	\frac{v_H^2\varphi_S^2}{v_S}\right) \nonumber\\
&+\frac{\lambda_{HS}}{4}(\varphi_H^2\varphi_S^2-\varphi_H^2v_S^2
-v_H^2\varphi_S^2) 
	+\mu_S^3\left(\varphi_S-\frac{1}{2}\frac{\varphi_S^2}{v_S}\right) 
	\nonumber\\
&+\frac{\mu'_S}{3}
	\left(\varphi_S^3-\frac{3}{2}v_S\varphi_S^2\right)+
	\frac{\lambda_S}{4}(\varphi_S^4-2v_S^2\varphi_S^2),
\end{align}
where $\varphi_H$ and $\varphi_S$ are constant background fields.
To avoid the potential unbounded from below, we impose
\begin{align}
\lambda_H>0 , \quad \lambda_S>0 , 
\quad \lambda_{HS}^2<4\lambda_H\lambda_S , 
\end{align}
where the last condition is needed for $\lambda_{HS}<0$.

Unlike the SM, there is a possibility that $V_0(\varphi_H, \varphi_S)$ has 
a global minimum which is different from the prescribed vacuum $(v_H, v_S)$.
Following the Refs. \cite{Funakubo:2005pu,Cheung:2010ba}, we define 
the various vacua as follows: 
\begin{align}
{\rm EW}&: v_H=246~{\rm GeV},\quad v_S=v_S^{\rm in}, \\
{\rm SYM}&: v_H=v_S=0, \\
{\rm I}&: v_H=0,\quad v_S\neq0, \\
{\rm II}&: v_H\neq0,\quad v_S=0, \\
{\rm III}&: v_H\neq 246~{\rm GeV},\quad v_S\neq v_S^{\rm in},
\end{align}
where $v_S^{\rm in}$ is the prescribed $v_S$. In the phase III, although both 
$v_H$ and $v_S$  are nonzero, they are different from the prescribed vacuum 
(EW phase).  Those various vacua are found by solving the following equations
\begin{align}
\frac{\partial V_0}{\partial \varphi_H}\bigg|_{\varphi_H=\bar{v}_H} 
&= \bar{v}_H\left[\lambda_H\bar{v}_H^2+\mu_{HS}\bar{v}_S
	+\frac{\lambda_{HS}}{2}\bar{v}_S^2-\mu_H^2 \right] = 0,\label{tree-tad_bv} \\
\frac{\partial V_0}{\partial \varphi_S}\bigg|_{\varphi_S=\bar{v}_S} 
&= \lambda_S\bar{v}_S^3+\mu'_S\bar{v}_S^2
	+\left(m_S^2+\frac{\lambda_{HS}}{2}\bar{v}_H^2\right)\bar{v}_S
	+\frac{\mu_{HS}}{2}\bar{v}_H^2+\mu_S^3 = 0.\label{tree-tad_bvS}
\end{align}
Note that one of the solutions corresponds to the EW phase.
For the EW phase to be the global minimum, we require 
\begin{align}
V_0(v_H,v_S) < V_0(\bar{v}_H\neq v_H, \bar{v}_S\neq v_S),\label{global_min}
\end{align}
where $\bar{v}_{H,S}$ denote the VEVs in the SYM, I, II and III
\footnote{
Since we will not consider a case in which both $\mu_{HS}$ and $\mu_S$ are simultaneously zero in the following discussion, the II phase would not be realized.
}.
To begin with, we demonstrate a comparison between the EW and I phases. 
The vacuum energies of the both phases are as follows.
\begin{align}
V_0^{\rm (EW)}(v_H, v_S)
&= -\frac{\lambda_H}{4}v_H^4-\frac{\mu_{HS}}{4}v_H^2v_S
	-\frac{\lambda_{HS}}{4}v_H^2v_S^2+\frac{\mu_S^3}{2}v_S-\frac{\mu'_S}{6}v_S^3
	-\frac{\lambda_S}{4}v_S^4, \\
V_0^{\rm (I)}(0, \bar{v}_S) &= \frac{\mu_S^3}{2}\bar{v}_S
	-\frac{\mu'_S}{6}\bar{v}_S^3-\frac{\lambda_S}{4}\bar{v}_S^4.
\end{align}
Here, we define $\Delta^{\rm (I-EW)} V_0$ by taking the difference of the two vacuum energies
\begin{align}
\lefteqn{\Delta^{\rm (I-EW)} V_0 \equiv V_0^{\rm (I)}(0, \bar{v}_S)-
V_0^{\rm (EW)}(v_H, v_S) }\nonumber\\
&=\frac{\lambda_H}{4}v_H^4+\frac{\mu_{HS}}{4}v_H^2v_S+
\frac{\lambda_{HS}}{4}v_H^2v_S^2
	+\frac{\mu_S^3}{2}(\bar{v}_S-v_S)-\frac{\mu'_S}{6}(\bar{v}_S^3-v_S^3)
	-\frac{\lambda_S}{4}(\bar{v}_S^4-v_S^4).
\end{align}
To satisfy the condition (\ref{global_min}), $\Delta^{\rm (I-EW)} V_0$ 
should be positive.  However, it could be negative if $\bar{v}_S$ gets large, 
which we will illustrate in the following.  For simplicity we take $\mu_S=0$. 
From Eq. (\ref{tree-tad_bvS}), it follows that
\begin{align}
\bar{v}_S = 0, \quad \frac{1}{2\lambda_S}
	\left[
		-\mu'_S\pm\sqrt{\mu'^2_S-4\lambda_S m_S^2}
	\right],
\end{align}
where the second solution corresponds to the I phase, 
and $\mu_S'^2\geq4\lambda_Sm_S^2$ should be satisfied for real solutions.
The vacuum energy of the I phase is reduced to
\begin{align}
V_0^{\rm (I)}(0, \bar{v}_S) = \frac{\bar{v}_S^2}{4}
	\left[	
m_S^2+\frac{-\mu'^2_S\pm \mu'_S\sqrt{\mu'^2_S-4\lambda_Sm_S^2}}{6\lambda_S}
	\right].
\end{align}
Therefore, for the large values of $\mu'_S$ and $m_S^2$ with their appropriate signs,
we may have $\Delta^{\rm (I-EW)} V_0<0$. 
As we will see later, the large $m_2$ can induce   such a case.

Now we consider another solution in Eq.~(\ref{tree-tad_bv}). 
The nonzero $\bar{v}_H$ is expressed as 
\begin{align} 
\bar{v}_H^2 = \frac{1}{\lambda_H}
\left[	
	-\mu_{HS}\bar{v}_S-\frac{\lambda_{HS}}{2}\bar{v}_S^2+\mu_H^2
\right],\label{eqn_barv_H}
\end{align}
where the real solution of $\bar{v}_H$ 
enforces $\mu_H^2>\lambda_{HS}\bar{v}_S^2/2+\mu_{HS}\bar{v}_S$.
Plugging this into Eq.~(\ref{tree-tad_bvS}), we have a cubic equation for $\bar{v}_S$.
If the cubic equation has only one real solution, it is nothing but $v_S$ and 
the III phase cannot exist.  On the other hand, if the cubic equation has other real solutions, 
and simultaneously Eq.~(\ref{eqn_barv_H}) has a real solution, the III phase would appear.
In such a case, it should be checked whether the energy level of the EW vacuum 
is lower than that of the III phase.
Let us denote the real solutions other than $v_S$ by $\bar{v}_S^{(1)}$ and $\bar{v}_S^{(2)}$.
We define $\bar{v}_S^{(2)}$ as the solution that gives a local maximum and thus
$\bar{v}_S^{(1)}$ and its corresponding solution of $\bar{v}_H$ yield the III phase.

Before going to the numerical analysis of the vacuum structures, 
we will obtain a range of $m_2$ consistent with the global vacuum conditions. 
To make the analysis simpler, we set $\alpha = \mu_S=0$. 
In the limit of $\sqrt{\lambda_{HS}}v_H\ll\sqrt{\lambda_S}|v_S|, m_2$, we may find
\begin{align}
\mu'_S \simeq -2\lambda_Sv_S+\frac{m_2^2}{v_S}, \quad
m_S^2 \simeq \lambda_Sv_S^2-m_2^2.\label{limiting_case}
\end{align}
The vacuum energy of the EW phase then can be cast into the form
\begin{align}
V_0^{\rm (EW)}(v_H, v_S)
= -\frac{\lambda_H}{4}v_H^4-\frac{\mu_S'}{6}v_S^3-\frac{\lambda_S}{4}v_S^4
\simeq \frac{\lambda_S}{12}v_S^4-\frac{1}{6}v_S^2m_2^2-\frac{\lambda_H}{4}v_H^4.
\end{align}
Requiring $\Delta^{\rm (SYM-EW)} V_0\equiv V_0^{\rm (SYM)}(0, 0)-V_0^{\rm (EW)}(v_H, v_S)>0$ 
yields the lower bound of $m_2$:
\begin{align}
\sqrt{\frac{\lambda_S}{2}}|v_S| < m_2.\label{m2_lower}
\end{align}
Furthermore, in the limit of $\sqrt{\lambda_S}|v_S|< m_2$,  
$\Delta^{\rm (I-EW)} V_0$ can be approximated as
\begin{align}
\Delta^{\rm (I-EW)} V_0\simeq -\frac{1}{12}\frac{(m_2^2)^4}{\lambda_S^3v_S^4}
	+\frac{1}{6}\frac{(m_2^2)^3}{\lambda_S^2v_S^2}.
\end{align}
Therefore, $\Delta^{\rm (I-EW)}V_0>0$ gives the upper bound of $m_2$:
\begin{align}
m_2 < \sqrt{2\lambda_S}|v_S|.\label{m2_upper}
\end{align}
Similarly, we may also obtain another constraint on $m_2$ from 
$\Delta^{\rm (III-EW)}V_0>0$.
Instead of doing so, we investigate the vacuum structures numerically.

\begin{figure}[t]
\center
\includegraphics[width=7cm]{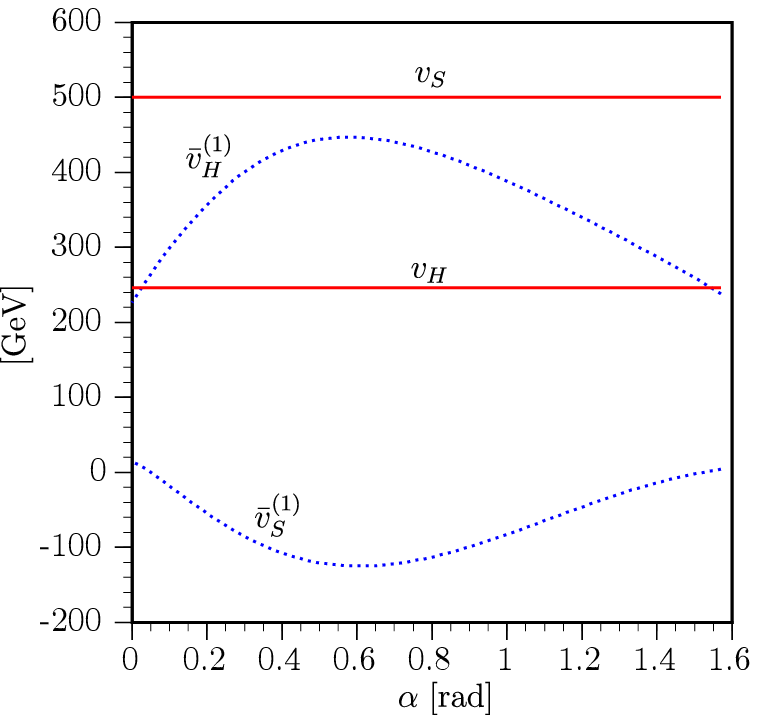}
\includegraphics[width=7.5cm]{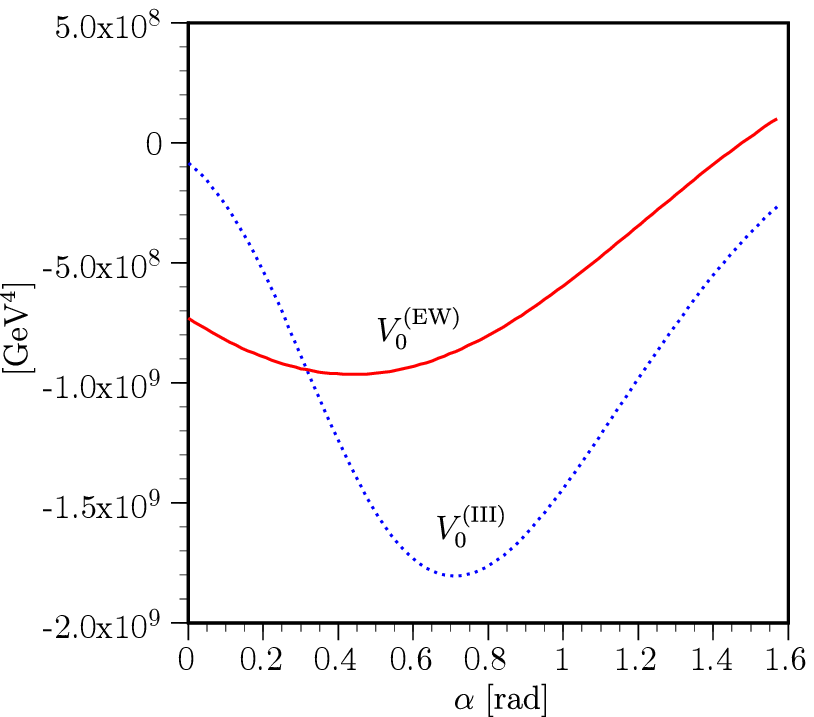}
\caption{(Left) The VEVs of the prescribed vacuum (EW phase) 
and the nontrivial vacuum (III phase) as a function of $\alpha$.
(Right) The vacuum energies of the two vacua as a function of $\alpha$,
where $V_0^{\rm (EW)}=V_0(v_H, v_S)$ and 
$V_0^{\rm (III)}=V_0(\bar{v}_H^{(1)}, \bar{v}_S^{(1)})$.
Here we take $m_1=125$ GeV, $m_2=200$ GeV, $v_S=500$ GeV, 
$\lambda_{HS}=0.01$, $\lambda_S=0.2$, $\mu_S=0$.}
\label{fig:vac_alpha}
\end{figure}

We begin with a case in which the EW vacuum becomes the local minimum
and the III phase can be the global minimum.
The representative example is shown in Fig.~\ref{fig:vac_alpha}.
We here take $m_1=125$ GeV, $m_2=200$ GeV, $v_S=500$ GeV, 
$\lambda_{HS}=0.01$,  $\lambda_S=0.2$.
In the left panel, the red lines represent the prescribed VEVs 
$(v_H, v_S)=$(246 GeV, 500 GeV)   and the upper and lower blue dotted curve denote 
$\bar{v}_H^{(1)}$ and $\bar{v}_S^{(1)}$  which significantly depend on the values of $\alpha$, 
and $\bar{v}_S^{(1)}$ is mostly negative.
In the right panel, the vacuum energies of the EW and III phases 
($V_0^{\rm (EW)}$ and$V_0^{\rm (III)}$) are shown, where $V_0^{\rm (III)}$ is given by
$V_0(\bar{v}_H^{(1)}, \bar{v}_S^{(1)})$.
We can see that $V_0^{\rm (EW)}<V_0^{\rm (III)}$ holds only up to $\alpha\simeq 0.3$ rad, 
and beyond this, the III phase becomes the global minimum.

\begin{figure}[t]
\center
\includegraphics[width=10cm]{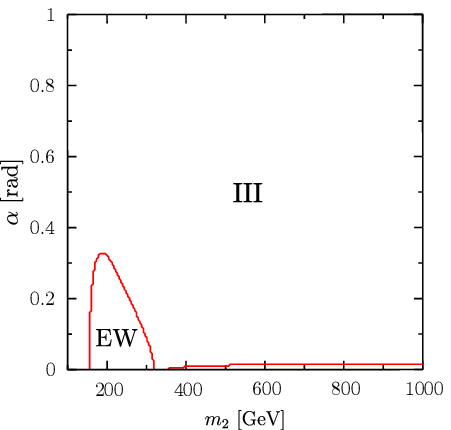} \\[0.5cm]
\includegraphics[width=7cm]{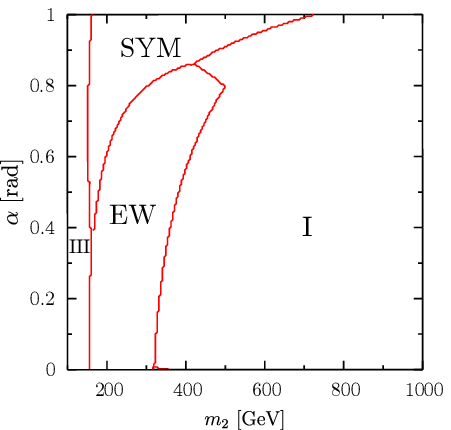}
\includegraphics[width=7cm]{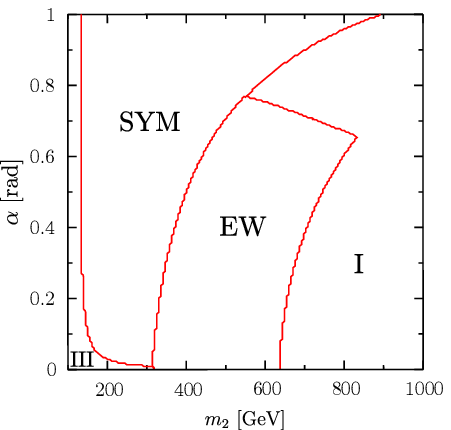}
\caption{The tree-level vacuum structures in the $\alpha$-$m_2$ plane.
(Upper) $v_S=500$ GeV; 
(Lower Left) $v_S=-500$ GeV; (Lower Right) $v_S=-1000$ GeV.}
\label{fig:vac_alpha_m2_tree}
\end{figure}
As suggested by Eqs.~(\ref{m2_lower}) and (\ref{m2_upper}), 
the region where the EW phase is the global minimum is also highly limited 
by the values of $m_2$  for a given $v_S$ and $\lambda_S$.
In Fig.~\ref{fig:vac_alpha_m2_tree} we illustrate such constraints.
In the upper panel, the vacuum structure is shown in the $\alpha$-$m_2$ plane,
taking the same input parameters as those in Fig.~\ref{fig:vac_alpha}.
It is found that the EW phase can be the global minimum only for 
160 GeV$\ltsim m_2\ltsim 320$ GeV,
and in most of the parameter space, the III phase is the global minimum. 
It should be emphasized that the possible range of $m_2$ around 
$\alpha\simeq 0$ rad is completely consistent with the analytic formulae 
$(\ref{m2_lower})$ and $(\ref{m2_upper})$.
From this observation, although we have not worked out the analytic formula 
from $\Delta^{\rm (III-EW)}V_0>0$, the mass bounds on the second Higgs boson
is more or less the same as $(\ref{m2_lower})$ and $(\ref{m2_upper})$.
In passing, we also find a region where the I phase becomes the global minimum 
at $\alpha\simeq0$ rad and $m_2\gtsim350$ GeV.

If the sign of $v_S$ is changed, namely, $v_S=-500$ GeV is taken but keeping 
the rest of the input parameters, the vacuum structure is drastically changed 
as shown in the left lower panel.  In this case, depending on $\alpha$ and $m_2$, 
the EW, SYM, I and III phases can become the global minimum.
From this plot, we obtain the upper bound of the heavy Higgs boson mass, i.e.,
$m_2\ltsim 500$ GeV around $\alpha=0.8$ rad.
Since $\alpha$ is relatively large, this upper bound cannot be obtained from 
$(\ref{m2_upper})$.

In the right panel, we set $v_S=-1000$ GeV.
We can see that the global minimum region of the EW phases is significantly affected 
by the value of $v_S$, rendering $m_2$ be as large as 800 GeV around 
$\alpha=0.65$ rad.   The possible lower value of $m_2$ is also pushed upward, 
$m_2\gtsim 300$ GeV.  For the small $\alpha$ region, the allowed range of $m_2$ 
nicely agrees with the analytic formulae $(\ref{m2_lower})$ and $(\ref{m2_upper})$.
%
%
\subsection{One-loop level analysis}
The tree-level Higgs potential receives the quantum corrections from the particles 
that couple to the Higgs fields. Therefore, the vacuum structure may change 
at the loop level.  In our model, the DM in the hidden sector can also affect the Higgs 
potential  through its couplings to the singlet Higgs field. So far, 
such an effect on the vacuum structure has not been investigated in the literature. 

We analyze the vacuum structure using the one-loop effective potential~\cite{Veff}
\begin{align}
V_1(\varphi_H, \varphi_S) &= \sum_{i}n_i
 \frac{\bar{m}_i^4}{64\pi^2}\left(\ln\frac{\bar{m}_i^2}{\mu^2}-c_i\right),
\end{align}
which is regularized in the $\overline{\rm MS}$ scheme;
$c=3/2$ for scalars and fermions and $c=5/6$ for gauge bosons. 
$\mu$ is a renormalization scale which will be set on $m_2$.
$\bar{m}$ is a field-dependent mass, and $i$ denotes particle species 
which is explicitly given by $i=H_{1,2}, G^0, G^\pm, W, Z, t,b,\psi$ 
and their degrees of freedoms ($n_i$) are respectively given by
\begin{align}
&n_{H_1}=n_{H_2}=n_{G^0} = 1,\quad n_{G^\pm} = 2,\quad n_W=6, \quad 
n_Z = 3, \nonumber \\  
& n_t=n_b=-12, \quad n_\psi=-4.
\end{align}
Unlike the tree-level analysis, it is impossible to obtain the analytic formulae 
of the vacuum structures.  We thus numerically minimize 
\begin{align}
V_{\rm eff}(\varphi_H, \varphi_S)=V_0(\varphi_H, \varphi_S)+
V_1(\varphi_H, \varphi_S)
\end{align}
and find a global minimum. Since we are considering the vacuum stability 
at the low energy  scale, we will concentrate on the fields space below 10 TeV.

We impose the one-loop tadpole conditions as
\begin{align}
\Vac{\frac{\partial V_{\rm eff}}{\partial \phi}} 
= \Vac{\frac{\partial V_0}{\partial \phi}}+\sum_in_i
	\frac{\bar{m}_i^2}{32\pi^2}\Vac{\frac{\partial \bar{m}_i^2}{\partial\phi}}
	\left(\ln\frac{m_i^2}{\mu^2}-c_i+\frac{1}{2}\right)=0,\label{1L-tad}
\end{align}
where $\phi=\varphi_H,\varphi_S$, 
and $\langle\cdots\rangle$ is defined such that a field-dependent quantity 
is evaluated in the EW vacuum, namely, $\varphi_H=v_H$ and $\varphi_S=v_S$.
Since we are taking $v_H$, $v_S$, $m_1$, $m_2$ and $\alpha$ as the input parameters
in place of $\mu_H^2$, $m_S^2$, $\lambda_H$, $\mu_{HS}$ and $\mu'_S$, 
we solve the following coupled equations numerically
\begin{align} \label{1loop-tadpole}
\frac{1}{v_H}\Vac{\frac{\partial V_{\rm eff}}{\partial \varphi_H}} 
=\frac{1}{v_S}\Vac{\frac{\partial V_{\rm eff}}{\partial \varphi_S}}&=0, \\
m_{hh}^2-m_1^2\cos^2\alpha-m_2^2\sin^2\alpha &=0,\\
m_{ss}^2-m_1^2\sin^2\alpha-m_2^2\cos^2\alpha &=0,\\
m_{hs}^2+(m_1^2-m_2^2)\sin\alpha\cos\alpha &=0,
\label{1loop-mass-mixing}
\end{align}
where $m_{hh}^2,~m_{ss}^2$ and $m_{hs}^2$ are defined by the mass matrix 
of the Higgs bosons at the one-loop level. 
The explicit expressions of one-loop quantities are listed in Appendix~\ref{app:1LHiggsMass}.
Since the treatment of the Nambu-Goldstone (NG)
boson loop contributions  are somewhat tricky and numerically unimportant, 
we will not take them into account in the analysis here.

\begin{figure}[t]
\center
\includegraphics[width=10cm]{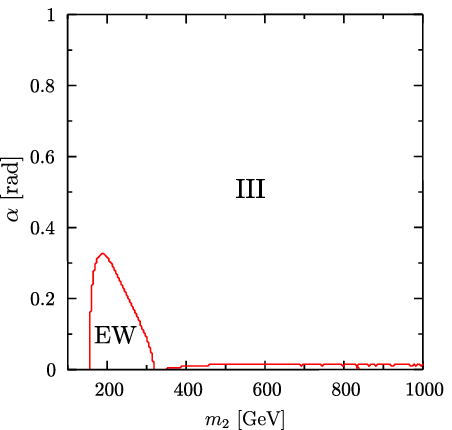} \\[0.5cm]
\includegraphics[width=7cm]{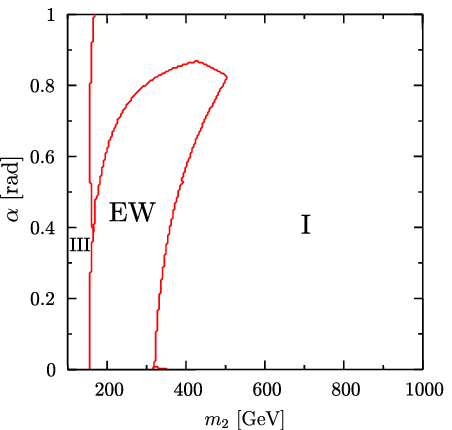}
\includegraphics[width=7cm]{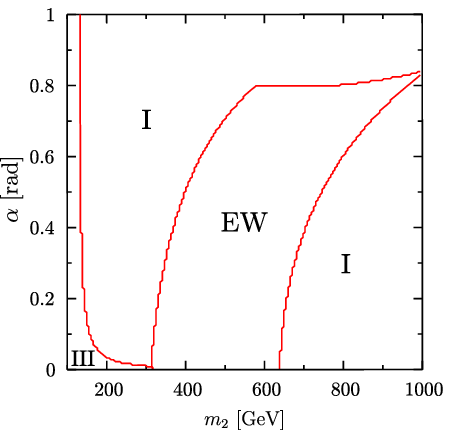} 
\caption{The one-loop level vacuum structures in the $\alpha$-$m_2$ plane.
(Upper) $v_S=500$ GeV; 
(Lower Left) $v_S=-500$ GeV; (Lower Right) $v_S=-1000$ GeV.
We take $m_\psi=\sqrt{\lambda_S}|v_S|/2$.}
\label{fig:vac_alpha_m2_1L}
\end{figure}

In Fig.~\ref{fig:vac_alpha_m2_1L}, the vacuum structures at the one-loop level 
are shown.  The input parameters in the Higgs sector are the same as in 
Fig.~\ref{fig:vac_alpha_m2_tree}. 
As for the parameters in the DM sector, as an example, we set
\begin{align}
m_{\psi_0} = \left(\frac{\sqrt{\lambda_S}}{2}-{\rm sgn}(v_S)\lambda\right)|v_S|, 
\quad \lambda = 0.2.
\end{align}
In the upper plot, the significant difference between the tree and one-loop 
results is not observed in the entire region.  As for the lower two plots, 
on the other hand, the region of the SYM phase appearing in the tree-level analysis 
vanishes, and the I phase region is enlarged instead. 
The reason is the following.
At the tree level, $(v_H, v_S)=(0,0)$ solution can exist if $\mu_S=0$ as can be seen
from Eqs.~(\ref{tree-tad_bv}) and (\ref{tree-tad_bvS}). 
At the one-loop level, on the contrary, there remain the constant terms in the tadpole condition
for $\varphi_S$ even after taking $v_H=0$ as we can observe in Eq.~(\ref{1L-tad_bvS}).
Those constant terms are proportional to $\mu_{HS}$
or $m_{\psi_0}$ which are always nonzero for the input parameters we are choosing here. 
Therefore, the SYM phase can never be realized at the one-loop level. 

For $v_S=-500$ GeV, the EW phase region is virtually unchanged 
and thus the allowed range of $m_2$ remains the same.
For $v_S=-1000$ GeV, on the other hand, although the vacuum structure in
the small $\alpha$ region is not much affected by the one-loop contributions,
the EW phase region is significantly distorted in $0.6~{\rm rad} <\alpha<0.8~{\rm rad}$, 
and the maximal $m_2$ can reach 1000 GeV around $\alpha=0.8$ rad.

Here, we remark that since the chosen $m_\psi$ can be the half of $m_2$ 
in the allowed region,  the DM relic density can be explained by the resonance effect 
as demonstrated in the previous section.

\begin{figure}[t]
\center
\includegraphics[width=7cm]{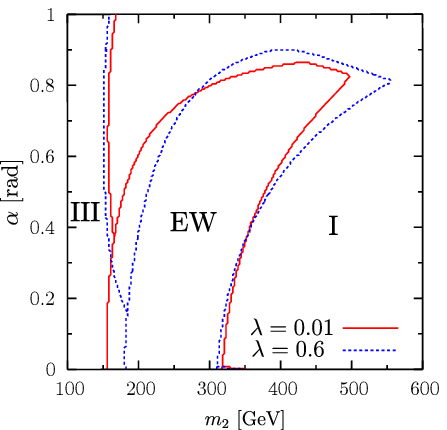}
\includegraphics[width=7cm]{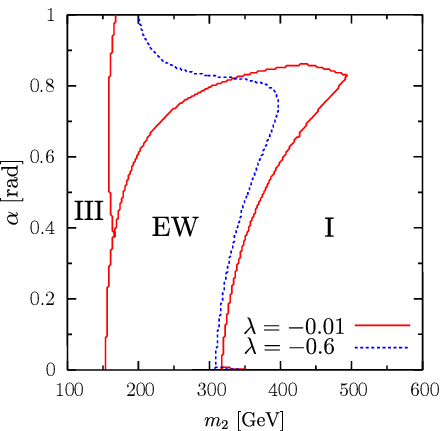} \\[0.5cm]
\includegraphics[width=7cm]{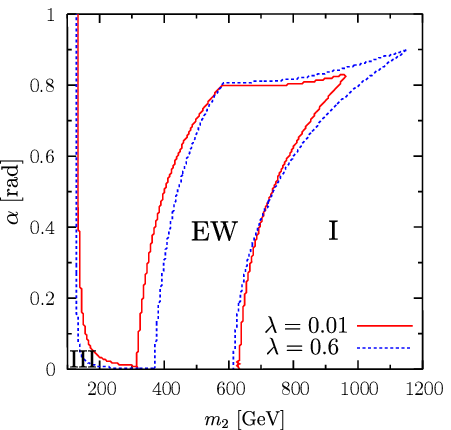}
\includegraphics[width=7cm]{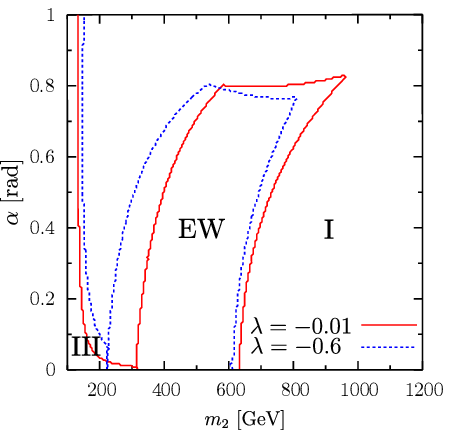} 
\caption{The effects of $\lambda$ on the vacuum structures. 
The red straight curves corresponds to the case of $|\lambda|=0.01$
and the blue dotted curves denotes to the case of $|\lambda|=0.6$.    
(Upper) $v_S=-500$ GeV; $\lambda=0.01, 0.6$ (left) and 
$\lambda=-0.01, -0.6$ (right).
(Lower) $v_S=-1000$ GeV; $\lambda=0.01, 0.6$ (left) and 
$\lambda=-0.01, -0.6$ (right).}
\label{fig:vac_alpha_m2_1L_DM}
\end{figure}

In order to see the DM loop effects on the vacuum structure, 
we vary $\lambda$ as
\begin{align}
\lambda = \pm 0.01,~\pm 0.6,
\label{DM_lam}
\end{align}
taking $m_{\psi_0}=|v_S|$ and the rest of the parameters are the same 
as the previous cases.  For the moment, we do not take account of the DM 
relic density constraint.  Our findings are shown in Fig.~\ref{fig:vac_alpha_m2_1L_DM}. 
In the upper (lower) panels, $v_S=-500~(-1000)$ GeV is taken,
and in the left (right) panels, we set $\lambda = 0.01, 0.6,~(-0.01,-0.6)$, 
where the red curves correspond to $\lambda=\pm 0.01$ and the blue dotted curves 
represent  $\lambda=\pm0.6$.  For the positive $\lambda$, 
the primary effect due to the increase of $\lambda$ is the shift of the EW phase 
region to the right side,  correspondingly, the minimal values of $m_2$ gets enhanced 
by about 25~(50) GeV  around $\alpha\simeq 0$ rad, and the maximal values of 
$m_2$ is enhanced by about 50~(200) GeV around $\alpha\simeq 0.8$ rad in the case 
of $v_S=-500~(-1000)$ GeV.

For the negative $\lambda$, on the other hand,  the EW phase region is more sensitive 
to the change of $\lambda$, especially $v_S=-500$ GeV case as shown in the upper 
right panel.  The III phase region which can exist in the $\lambda=-0.01$ case is gone 
if $\lambda=-0.6$, and the EW phase region is much enlarged. 
In this case, the range of $m_1<m_2\ltsim 200$ GeV is allowed for any $\alpha\in [0, 1]$. 
The maximal value of $m_2$, however, is reduced to 400 GeV which is realized at 
$\alpha\simeq 0.75$ rad.

In the case of $v_S=-1000$ GeV, the minimal value of $m_2$ is increased by about 
100 GeV, and the maximal value of it gets bigger by about 150 GeV  when $\lambda$ is 
changed from $-0.6$ to $-0.01$.

Here, we comment on the DM relic density constraint. The choice of $m_{\psi_0}=|v_S|$
with Eq.~(\ref{DM_lam}) does not yield $m_\psi=m_2/2$ in the EW phase region.
Therefore, the right amount of the DM relic density 
may not be guaranteed. It turns out that the change of $\lambda$ has little effect 
on the vacuum structures once we take $m_\psi=m_2/2$. 

In summary, in this section we showed that the diverse types of the (false) vacua 
are realized in this model, which is due to the presence of the singlet scalar field. 
Most important consequence is that the EW vacuum is not always the global minimum
and often becomes the metastable state. 
As explicitly demonstrated here, the global minimum condition for the EW vacuum can 
eliminate the large portion of the parameter space.
As a result, we can obtain the strong bounds on $m_2$. For $\alpha\ltsim0.2$ rad, 
we find  $\sqrt{\lambda_S/2}|v_S|\ltsim m_2\ltsim \sqrt{2\lambda_S}|v_S|$.
Is should be stressed that this mass bound exclusively depends on $v_S$ and $\lambda_S$ 
and not on the value of $\lambda_{HS}$.
For $v_S\ltsim v_H$, however, the above $m_2$ range would not be valid any more.
The EW phase regions appear somewhere near $\alpha=\pi/2$ 
in which $m_2$ becomes the SM-like Higgs boson. 

At the loop level, the DM also contributes to the effective potential.
It is found that the vacuum structure has some sensitivity to the magnitude of $\lambda$,
rendering the viable ranges of $\alpha$ and $m_2$ changed.
However, such effects would be diminished once $m_\psi=m_2/2$ is imposed, 
which is indeed the case if we wish to saturate the observed thermal relic density of CDM 
by a singlet fermion CDM.

\section{Vacuum stability}\label{sec:vac_stabil}

Recent results of ATLAS and CMS experiments may indicate $m_h \sim 125 \GeV$.
If SM is assumed to be valid up to a very high energy scale, for example, GUT or Planck scale, the SM Higgs potential realizing such a light higgs faces the problem of vacuum instability, that is, the existence of non-SM deeper vacuum or unbound-from-below caused by the negativity of quartic self-coupling of Higgs at large field region.
However, a simple extension of SM changes this situation drastically.
As shown in the previous section, including a singlet which couples to SM via Higgs portal make vacuum structure very complicated.
In this section, we show how the SM picture of instability problem of Higgs potential at high energy scale is changed in our model.

\begin{figure}[htbp]
\begin{center}
\includegraphics[width=0.8\textwidth]{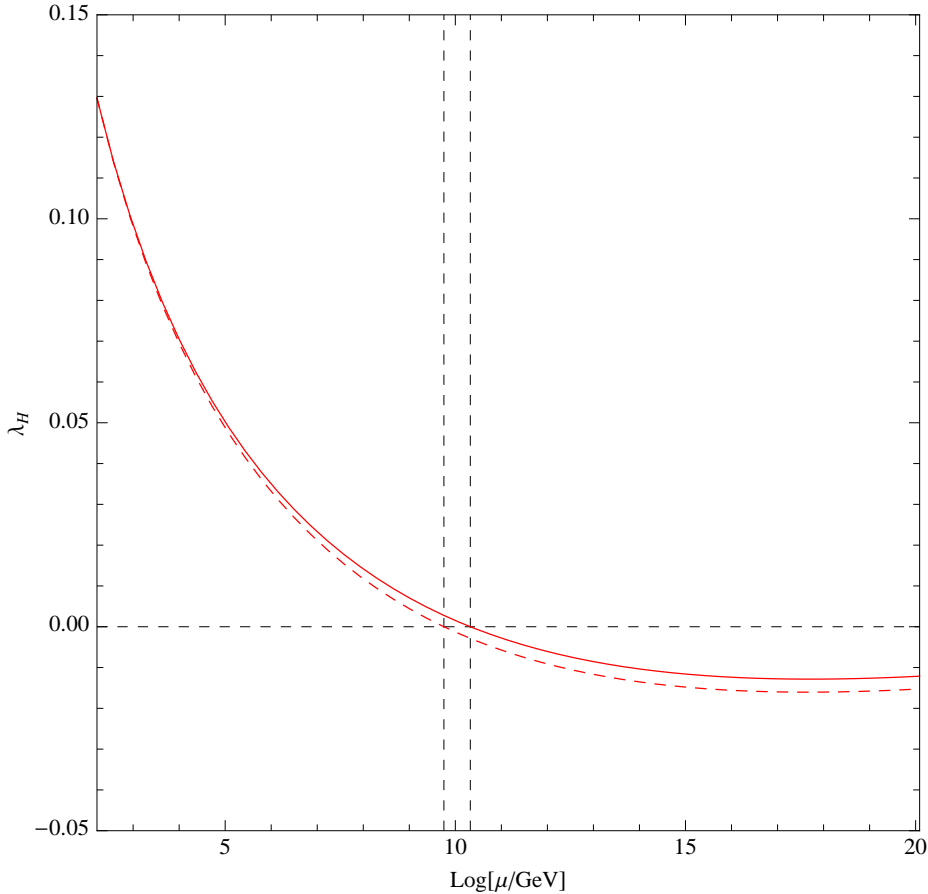}
\caption{The scale dependence of SM Higgs quartic coupling $\lambda_H$ for $m_h=125 \GeV$. 
The solid and dashed red lines correspond to the cases with $\overline{\rm MS}$ top-Yukawa couplings obtained at the matching scale $m_t(M_t) \simeq 163.5 \GeV$ and $m_t(m_t) \simeq 164 \GeV$, respectively.
The instability scale is different from each other by about $0.5 \GeV$.
}
\label{fig:lambdaH-SM}
\end{center}
\end{figure}

Before getting into the analysis, we fix the low energy boundary quantities at the scale of $Z$-boson pole mass.
\beq
\alpha_{\rm em}(M_Z) = \frac{1}{127.926} \quad , \quad \alpha_2(M_Z) = \frac{\alpha_{\rm em}(M_Z)}{\sin^2 \theta_W} \quad , \quad \alpha_3(M_Z) = 0.1184
\eeq
where 
\beq
M_Z = 91.188 \GeV \quad , \quad \sin^2 \theta_W = 0.2312 .
\eeq
We run up the couplings to the scale of top-quark running mass in $\overline{\rm MS}$-scheme.
Denoted as $m_t$, the running mass is obtained from the well-known formulas listed in Appendix.
For the top-quark pole-mass \cite{Lancaster:2011wr}
\beq
M_t = 173.2 \GeV ,  
\eeq
we find 
\beq
m_t(m_t) \simeq 164.0 \GeV .
\eeq
Correspondingly, the running top-Yukawa coupling is given by
\beq
\lambda_t(m_t) = \sqrt{2} m_t(m_t) / v .
\eeq
Note that choosing $M_t$ as the matching scale results in $m_t(M_t) \simeq 163.5 \GeV$ which is smaller than $m_t(m_t)$ by about $0.5 \GeV$.
As shown in Fig.~\ref{fig:lambdaH-SM}, this difference results in about $0.5 \GeV$ higher instability scale of SM Higgs potential.
It is similar size to the uncertainties of the top-quark pole mass and associated instability scale.  

In order to obtain the running Higgs quartic coupling at $m_t(m_t)$, we solved Eqs.~(\ref{1loop-tadpole})-(\ref{1loop-mass-mixing}) numerically, ignoring momentum dependent corrections which is about \%-level contributions \cite{Allanach:2004rh}
\footnote{
In case of SM, a \%-level correction to Higgs mass results in about an order of magnitude change of the instability scale.  
}.

\subsection{Tree-level (mixing) effect}
The first thing we should note in a extension of SM like ours is how the mass of the SM-like higgs is determined.
For the fixed VEV of SM Higgs field, $\lambda_H$ is no more the only parameter which determines the mass of SM-like higgs.
As studied in Refs. \cite{Lebedev:2012zw,EliasMiro:2012ay}, tree-level contribution of a singlet scalar can remove the instability problem in a very simple manner.
Similarly, the tree-level effect of our model on the stability of SM Higgs potential can be read off from the mixing effect on the SM Higgs.
At tree level, the SM Higgs quartic coupling can be written as
\beq \label{lambdaH-tree}
\lambda_H = \left[ 1 + \tan^2 (\alpha) \frac{m_2^2}{m_1^2} \right] \cos^2 (\alpha) \frac{m_1^2}{2 v^2}
\eeq 
where $\alpha$ is the mixing angle.
If $\alpha=0$(no mixing), one obtains 
$\lambda_H=\lambda_H^{\rm SM} \equiv m_h^2 /(2 v^2)$ with
$m_h = 125 \GeV$.
However, once the mixing is turned on, $\lambda_H$ can be much larger than 
$\lambda_H^{\rm SM}$.  

It may turn out that Higgs is very SM-like, so the mixing between the singlet and 
the SM Higgs might have to be highly suppressed. 
Even in such a case, we can still have a sizable increase of $\lambda_H$ to achieve
the stability of Higgs potential by pushing up $m_2$ to its unitary bound at most.
If small, the mixing angle is  given as
\beq
\alpha \simeq \l| \frac{m_{hs}^2}{m_{ss}^2 - m_{hh}^2} \r|
= \l| \frac{v_H \l( \mu_{HS} + \lambda_{HS} v_S \r)}{m_{ss}^2 - m_{hh}^2} \r| .
\eeq
In the limit of large $v_S$ where the heavy singlet-like Higgs is likely to be decoupled, 
$\lambda_H$ is approximated as 
\beq
\lambda_H \simeq \lambda_H^{\rm SM} + \frac{1}{4} \frac{\lambda_{HS}^2}{\lambda_S}
\eeq
which is the same as the case of Ref. \cite{EliasMiro:2012ay,Lebedev:2012zw}.
On the other hand, if $\mu_{HS} \approx - \lambda_{HS} v_S$, the tree-level effect on $\lambda_H$ is negligible, but the loop-effect from extra particles could be still large enough to remove the vacuum instability as long as $\lambda_{HS}$ is sizable.

\begin{figure}[htbp]
\begin{center}
\includegraphics[width=0.8\textwidth]{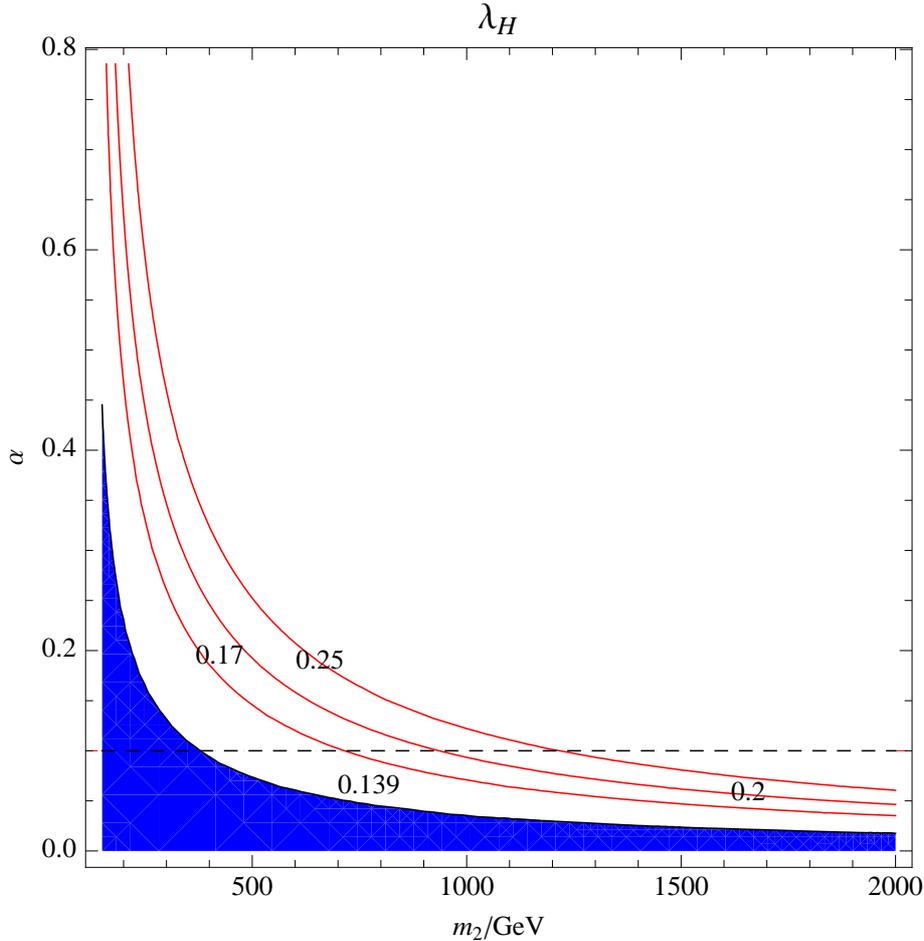}
\caption{Tree-level mixing effect on $\lambda_H$ for $m_1=125 \GeV$ in $(m_2, \alpha)$ plane. 
The $x$-axis ($\alpha=0$) corresponds to pure SM where $\lambda_H(m_t) = 0.1277$.
In SM, vacuum stability up to Planck scale requires $\lambda_H(m_t) \gtrsim 0.139$(white region) corresponding to $m_h\simeq130\GeV$ for $M_t=173.2 \GeV$ and $\alpha_s(M_Z)=0.1184$.
The dashed black-line corresponding to $\alpha=0.1$ is a reference line that might be imposed from LHC Higgs searches.
}
\label{fig:VS-tree}
\end{center}
\end{figure}
%
In Fig. \ref{fig:VS-tree}, we show $\lambda_H$ contours in ($m_2$,$\alpha$)-plane.
If the RG-running of $\lambda_H$ is pure SM-like, vacuum stability requires $m_h \gtrsim 130 \GeV$ (corresponding to $\lambda_H \gtrsim 0.139$) at 2-loop level.
We can see in the figure that, even if $\alpha \lesssim 0.1$, vacuum can be stable if $m_2 \gtrsim 400 \GeV$.

\subsection{Loop effect}
\subsubsection{RG equations}
If $\alpha$ is limited to be small due to constraints from collider experiments and/or direct searches of dark matter,  we may have to resort to the additional loop contributions coming from the Yukawa couplings, $\lambda_{HS}$, $\lambda_S$ and $\lambda$.

The $\beta$-function of a coupling $\lambda_i$ in the renormalization group equation is defined as
\beq
\beta_{\lambda_i} \equiv d \lambda / d \ln \mu
\eeq
where $\mu$ is the renormalization scale.
For SM gauge couplings, the $\beta$-functions are given as
\beq
\beta_{g_a} = \frac{1}{16 \pi^2} b_a^{(1)} g_a^3 + \frac{1}{\left( 16 \pi^2 \right)^2} b_a^{(2)} g_a^3
\eeq
where, for notational convenience, we have redefined the SM couplings as 
\beq
g_1 \equiv g' \ , \ g_2 \equiv g \ , \ g_3 \equiv g_s
\eeq
and
\bea
b_a^{(1)} &=& \l(\frac{41}{6} , - \frac{19}{6}, -7 \r) ,
\\
b_a^{(2)} &=& \sum_b c_{ab} g_b^2 - d_a \lambda_t^2
\eea
with
\beq
c_{ab} =\l(
\begin{array}{ccc}
199/18 & 9/2 & 44/3
\\
3/2 & 35/6 & 12 
\\
11/6 & 9/2 & -26
\end{array}
\r)
\ , \
d_a = \l(\frac{17}{6}, \frac{3}{2}, 2 \r) .
\eeq
For dimensionless couplings (including the top Yukawa coupling) in the scalar potential, the 1-loop $\beta$-functions are as follows. 
\bea \label{lambdat-1loop-beta}
\beta_{\lambda_t}^{(1)} &\simeq& \frac{1}{16 \pi^2} \lambda_t \l[ \frac{9}{2} \lambda_t^2 - \l( 8 g_3^2 + \frac{9}{4} g_2^2 + \frac{17}{12} g_1^2 \r) \r] ,
\\
\label{lambdaH-1loop-beta}
\beta_{\lambda_H}^{(1)}&=& \frac{1}{16 \pi^2} \l[ 24 \lambda_H^2 + 12 \lambda_H \lambda_t^2 - 6 \lambda_t^4 - 3 \lambda_H \l( 3 g_2^2 + g_1^2 \r) + \frac{3}{8} \l( 2 g_2^4 + \l( g_2^2 + g_1^2 \r)^2 \r) + \frac{1}{2} \lambda_{HS}^2 \r] ,~~~~~~~
\\
\label{lambdaHS-1loop-beta}
\beta_{\lambda_{HS}}^{(1)} &=& \frac{\lambda_{HS}}{16 \pi^2} \l[ 2 \l( 6 \lambda_H + 3 \lambda_S + 2 \lambda_{HS} \r) - \l( \frac{3}{2} \lambda_H \l( 3 g_2^2 + g_1^2 \r) - 6 \lambda_t^2 - 4 \lambda^2 \r) \r] ,
\\
\label{lambdaS-1loop-beta}
\beta_{\lambda_S}^{(1)} &=& \frac{1}{16 \pi^2} \l[ 2 \lambda_{HS}^2 + 18 \lambda_S^2 + 8 \lambda_S \lambda^2 -  8 \lambda^4 \r] ,
\\
\label{lambda-1loop-beta}
\beta_\lambda^{(1)} &=& \frac{5}{16 \pi^2} \lambda^3 .
\eea
For the top-Yukawa and quartic self couplings of Higgs field, the 2-loop $\beta$-function contributions are as follows.
\bea
\beta_{\lambda_t}^{(2)} &\simeq& \frac{1}{\l( 16 \pi^2 \r)^2} \lambda_t \left[ -12 \lambda_t^4 - 12 \lambda_t^2 \lambda_H + 6 \lambda_H^2 + \lambda_t^2 \l( 36 g_3^2 + \frac{225}{16} g_2^2 + \frac{131}{16} g_1^2 \r) \right.
\nonumber \\
&& \left. + g_3^2 \l( 9 g_2^2 + \frac{19}{9} g_1^2 \r) - \frac{3}{4} g_2^2 g_1^2 - 108 g_3^4 - \frac{23}{4} g_2^4 + \frac{1187}{216} g_1^4
\right] ,
\\
\beta_{\lambda_H}^{(2)} 
&\simeq& 
\frac{1}{\l( 16 \pi^2 \r)^2} \l[ \lambda_H \lambda_t^2 \l( \frac{85}{6} g_1^2 + \frac{45}{2} g_2^2 + 80 g_3^2 - 144 \lambda_H - 3 \lambda_t^2 \r) \right.
\nonumber \\
&& \left. + \lambda_H \l( \frac{629}{24}g_1^4 - \frac{73}{8} g_2^4 + \frac{39}{4} g_1^2 g_2^2 + \l( 36 g_1^2 + 108 g_2^2 \r) \lambda_H - 312 \lambda_H^2 \r) \right.
\nonumber \\
&& \left. + \lambda_t^2 \l( - \frac{19}{4} g_1^4 - \frac{9}{4} g_2^4 + \frac{21}{2} g_1^2 g_2^2 - \l( \frac{8}{3} g_1^2 + 32 g_3^2 \r) \lambda_t^2 + 30 \lambda_t^4 \r) \right.
\nonumber \\
&& \left. + \frac{1}{48} \l( 915 g_2^6 - 379 g_1^6 - 289 g_1^2 g_2^4 - 559 g_1^4 g_2^2 \r) \r] .
\eea
Note that $\beta_{\lambda_{HS}}$ is linearly proportional to $\lambda_{HS}$, hence $\lambda_{HS}$ does not change its sign during its RG-running.
$\lambda$ does not change its sign, too, and it increases or decreases monotonically, depending its sign.
$\lambda$ appears in the RGEs of $\lambda_{HS}$ and $\lambda_S$ as a squared one, so its sign does not affect RG-running of those couplings. 

As can be seen from \eq{lambda-1loop-beta}, the RGE of $\lambda$ does not depend on any other couplings at 1-loop level.
Hence it can be solved easily, giving the solution 
\beq \label{lambda-run}
\lambda(\mu) = \l[ \frac{1}{\lambda^2(\mu_0)} - \frac{5}{8 \pi^2} \ln \frac{\mu}{\mu_0} \r]^{-1/2} .
\eeq
We find that
$\beta_\lambda \leq 1$ if 
\beq
\lambda(M_\pl) \leq \l( \frac{16 \pi^2}{5} \r)^{1/3}
\eeq
which corresponds to 
\beq
\lambda(m_t) = \l[ \frac{1}{\lambda(M_\pl)^2} + \frac{5}{8 \pi^2} \ln \l( \frac{M_\pl}{m_t} \r) \r]^{-1/2} \leq 0.625
\eeq
where $M_\pl \simeq 1.2 \times 10^{19} \GeV$ is the Planck mass.
Note that $\beta_{\lambda_S}$ has a strong dependence on $\lambda$, hence as $\lambda(\mu_0)$ becomes close to upper bound, the allowed band of $\lambda_S(\mu_0)$ for a perturbative positive $\lambda_S(\mu)$ becomes narrower and eventually disappears.

\subsection{Numerical analysis}
The aim of our numerical analysis is to see if the demand of vacuum stability constrains our model parameters which should satisfy EWPT, DM relic density and DM direction detection bound, and if there is any way to probe the model in future collider experiments.
In this regard, the crucial parameters are 
\beq
m_1, \quad m_2, \quad \alpha, \quad \lambda, \quad \lambda_H, \quad \lambda_{HS}, \quad \lambda_S .
\eeq
These parameters are involved in the following constraints.
\begin{itemize}
\item EWPT: $m_1$, $m_2$ and $\alpha$
\item DM relic density: The first 4 parameters
\item DM direct searches: The first 4 parameters
\item Vacuum stability ($\lambda_H(\mu) > 0$ and $\lambda_S(\mu) > 0$): All of them.  
\end{itemize}
Note that not all of those parameters are independent.
For example, if we choose $m_1$, $m_2$ and $\alpha$ as an input, each element of the mass matrix is fixed
through Eq.~(\ref{conversion}).
Since $v_H = 246 \GeV$, $\lambda_H$ is fixed by the first line of Eq.~(\ref{lam_trade}), but we are free to choose $\lambda_{HS}$ and $\lambda_S$.
Once ($\lambda_{HS}, \lambda_S$) is chosen, ($\mu_{HS}, \mu_S'$) is given in terms of $v_S$ by the second and third lines of \eq{lam_trade}, respectively, where we assume $\mu_S = 0$.
Therefore, the free parameters we can use in analyzing the scale dependence of dimensionless couplings are
\beq
m_1, \quad m_2, \quad \alpha, \quad \lambda, \quad \lambda_{HS}, \quad \lambda_S .
\eeq

Inspired by the recent results of LHC experiments, we take $m_1 = 125 \GeV$ and vary $m_2$ and $\alpha$ in the ranges,
\beq
150 \GeV \leq m_2 \leq 2 \TeV, \quad  0 \leq \alpha \leq \pi/4 .
\eeq
Since we are interested in parameter space where couplings do not blow up, we consider $\lambda$ ranging  
\beq \label{lambda-band}
0.01 \leq \lambda \leq 0.6 .
\eeq
We find that $\lambda_{HS}$ blows up if $\lambda_{HS}(m_t) \gtrsim 0.4$ even for $\lambda_H(m_t) = \lambda_H^{SM}(m_t)$ and $\lambda_S(m_t) = \lambda(m_t) = 0$.
$\lambda_S$ blows up if $\lambda_S(m_t) \gtrsim 0.26$ for $\lambda(m_t) = 0.6$ and $\lambda_{HS}(m_t)  = 0$.
If $\lambda_{HS}$ starts from a negative value, $\lambda_S$ blows up if $\lambda_{HS}(m_t) \lesssim -0.9$ for $\lambda(m_t) \lesssim 0.6$ and $\lambda_S(m_t) \lesssim 0.26$.
Based on this observation, we scan the following ranges of our parameters.
\beq
-0.9 \leq \lambda_{HS}(m_t) \leq 0.4, \quad 0.01 \leq \lambda_S(m_t) \leq 0.26 .
\eeq

\begin{figure}[htbp]
\begin{center}
\includegraphics[width=0.45\textwidth]{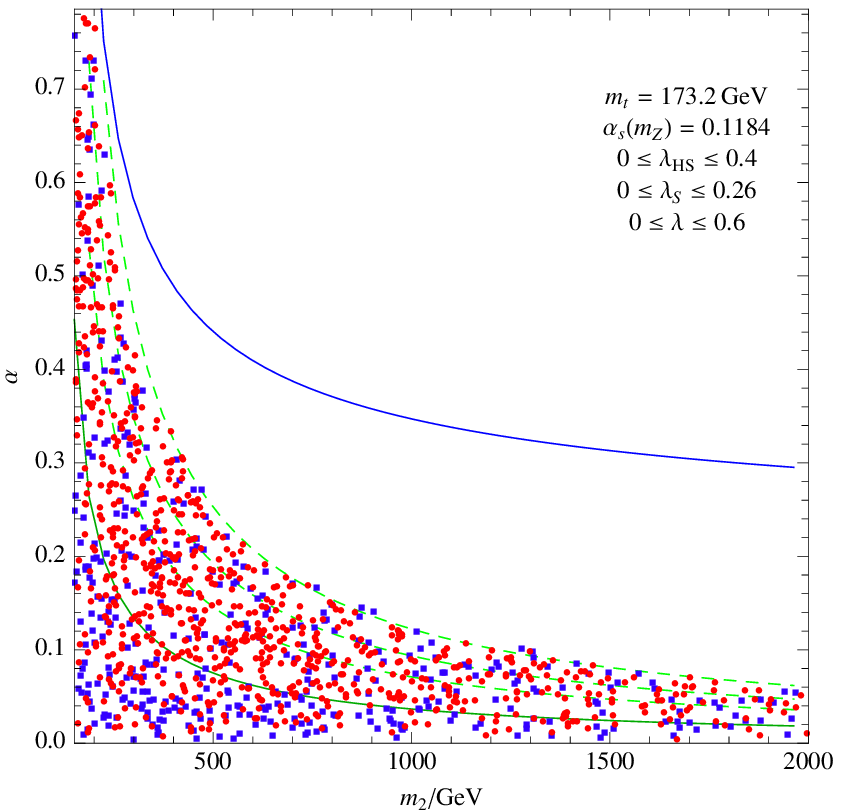}
\includegraphics[width=0.45\textwidth]{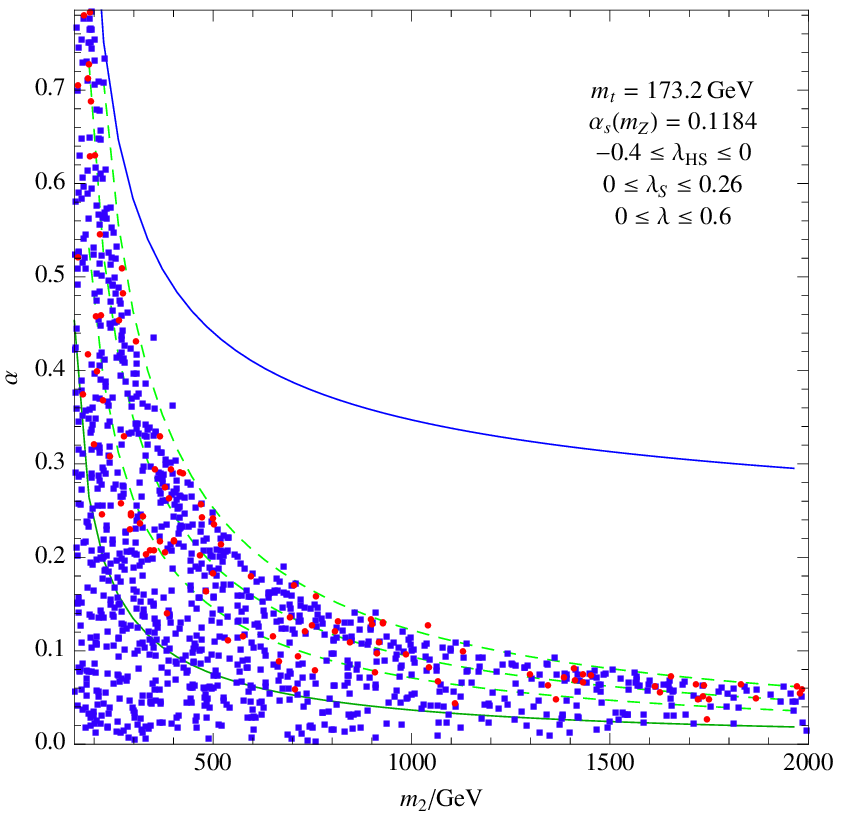}
\caption{Vacuum stability in ($m_2$,$\alpha$)-plane.
Left: $\lambda_{HS} \geq 0$.
Right: $\lambda_{HS} < 0$.
Blue line is the EWPT bound.
Green solid line is the bound for stable vacuum when RG-run is SM-like.
Green dashed lines correspond to $\lambda_H = 0.17, 0.20, 0.25$ from bottom to top.
Red/blue dots indicate stable/unstable vacuum.}
\label{fig:VS-2loop}
\end{center}
\end{figure}

\begin{figure}[htbp]
\begin{center}
\includegraphics[width=0.45\textwidth]{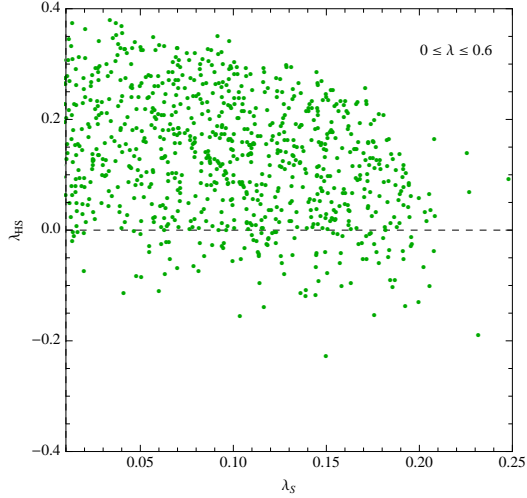}
\caption{The distribution of stable vacua appeared in Fig.~\ref{fig:VS-2loop} in ($\lambda_S$,$\lambda_{HS}$)-plane.}
\label{fig:lambdaHS-vs-lambdaS}
\end{center}
\end{figure}

\begin{figure}[htbp]
\begin{center}
\includegraphics[width=0.45\textwidth]{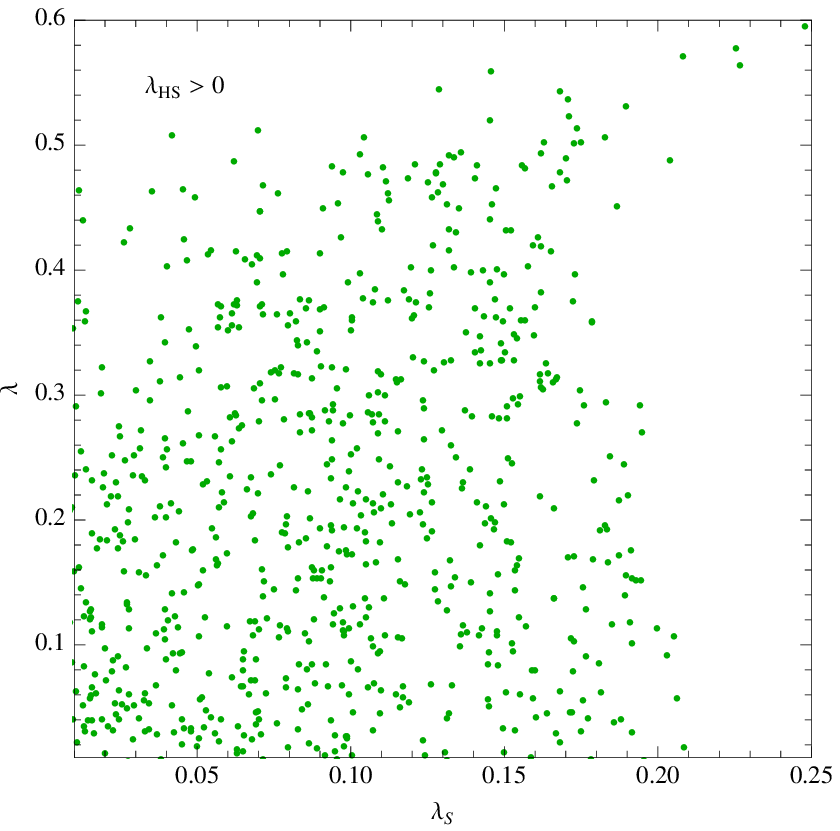}
\includegraphics[width=0.45\textwidth]{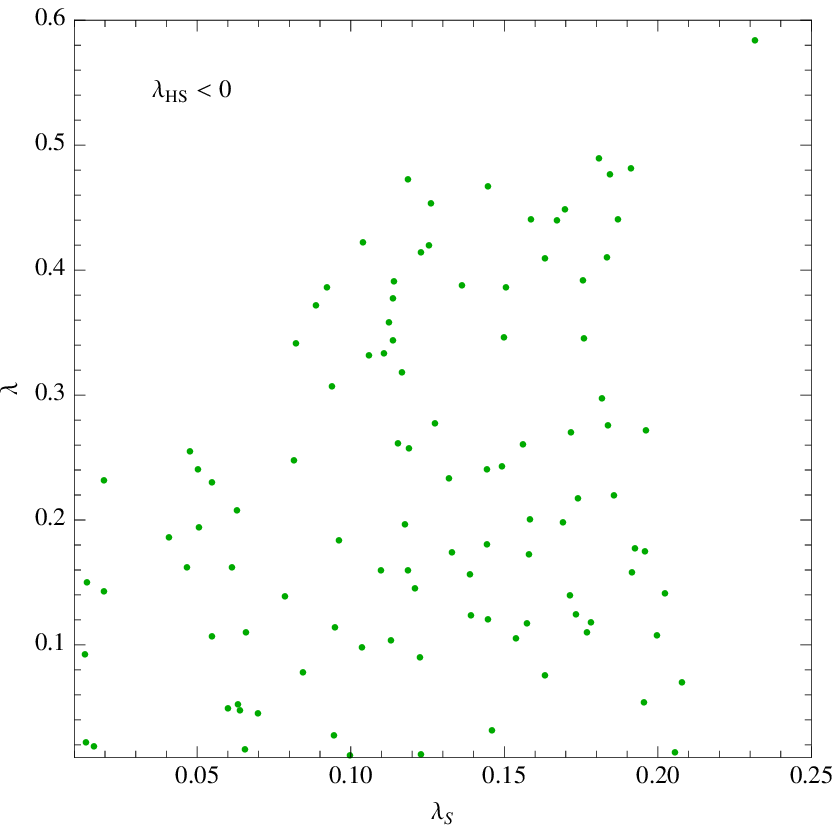}
\caption{The distribution of stable vacua appeared in Fig.~\ref{fig:VS-2loop} in ($\lambda_S$,$\lambda$)-plane.
Left: $\lambda_{HS} \geq 0$.
Right: $\lambda_{HS} < 0$.}
\label{fig:lambda-vs-lambdaS}
\end{center}
\end{figure}

Fig. \ref{fig:VS-2loop} shows a distribution of stable(red dots)/unstable(blue dots) vacua  in ($m_2$, $\alpha$) plane for $10^4$ randomly chosen parameter sets.
Left and right panels are for positive and negative $\lambda_{HS}$, respectively.
In both panels, the upper bound of red/blue dots are from the perturbativity constraints which we chose 
\beq
\beta_i \leq 1 .
\eeq
In case of $\lambda_{HS}  \geq 0$(left panel),  red dots covers whole region below the perturbativity bound and go far below the SM bound.
They appear even in the region where mixing is quite small.
This is because non-zero $\lambda_{HS}$ can provide a large enough loop effect on the running of $\lambda_H$ to remove the vacuum instability even if it is upper-bounded to a rather small value to avoid blow-up.
On the other hand, for $\lambda_{HS} <0$(right panel), quite small number of red dots(stable vacua) appear.
This is because the constraint from unbounded-from-below at large field region removes big chunk of parameter space, which is clear from Fig.~\ref{fig:lambdaHS-vs-lambdaS} and~\ref{fig:lambda-vs-lambdaS} where distributions of stable vacua shown in Fig.~\ref{fig:VS-2loop} are depicted in ($\lambda_{HS}$,$\lambda_S$) and ($\lambda$,$\lambda_S$) planes, respectively.

\begin{figure}[htbp]
\begin{center}
\includegraphics[width=0.45\textwidth]{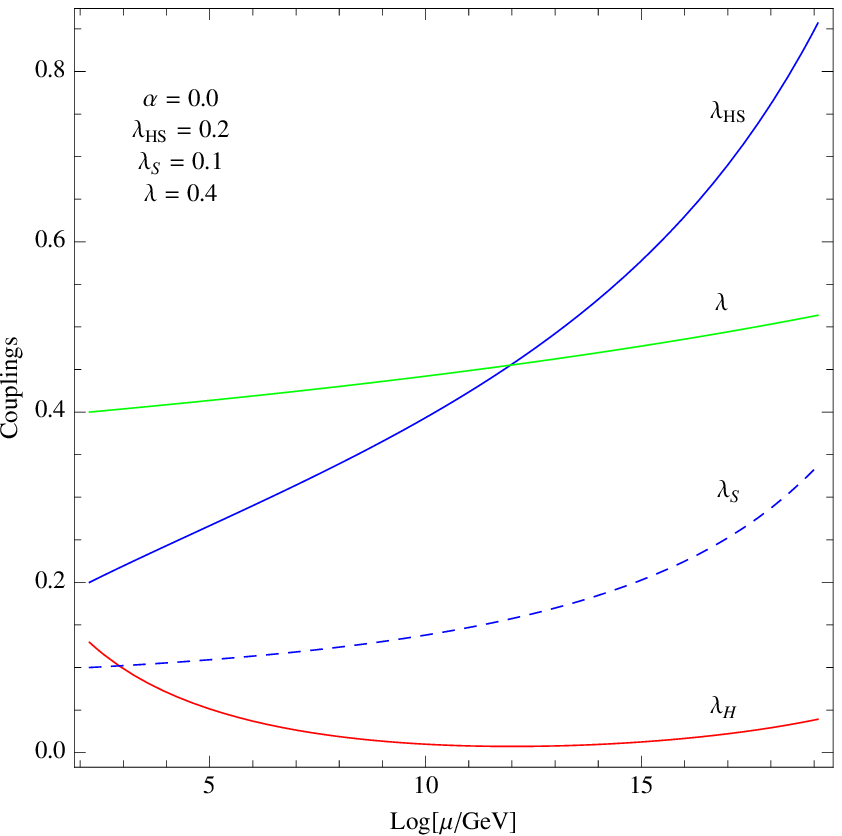}
\includegraphics[width=0.45\textwidth]{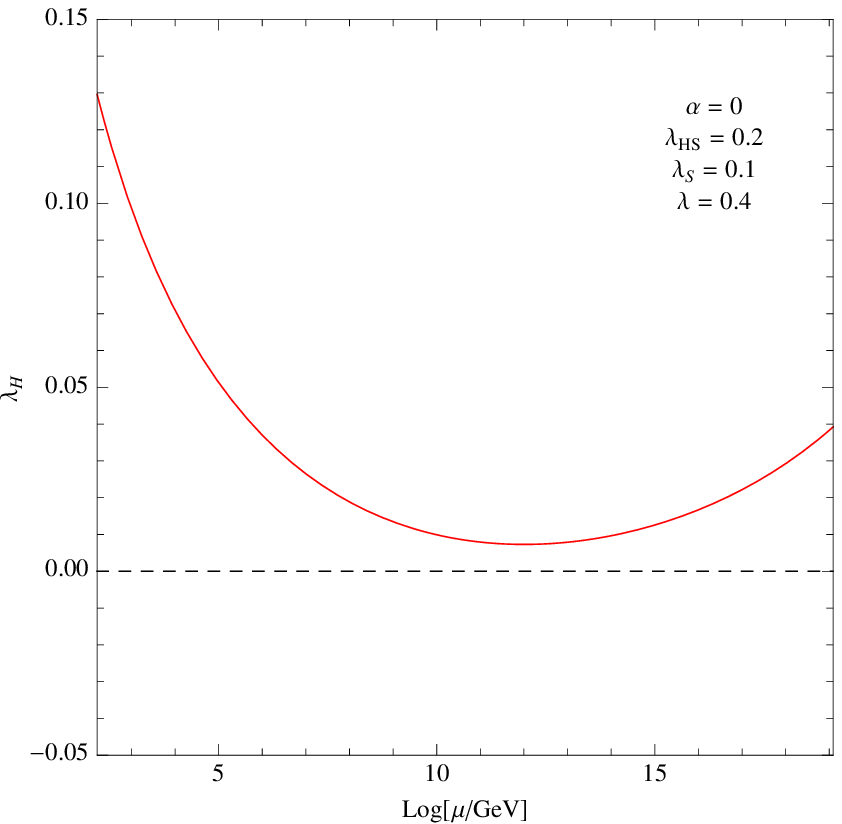}
\caption{RG-running of couplings as a function of renormalization scale for $m_1 = 125 \GeV$, $m_2=500 \GeV$ and $\alpha=0$, i.e., no-mixing.
Red/blue/green/dashed-blue line corresponds to $\lambda_H$/$\lambda_{HS}$/$\lambda$/$\lambda_S$.}
\label{fig:runnings-nomix}
\end{center}
\end{figure}

\begin{figure}[htbp]
\begin{center}
\includegraphics[width=0.45\textwidth]{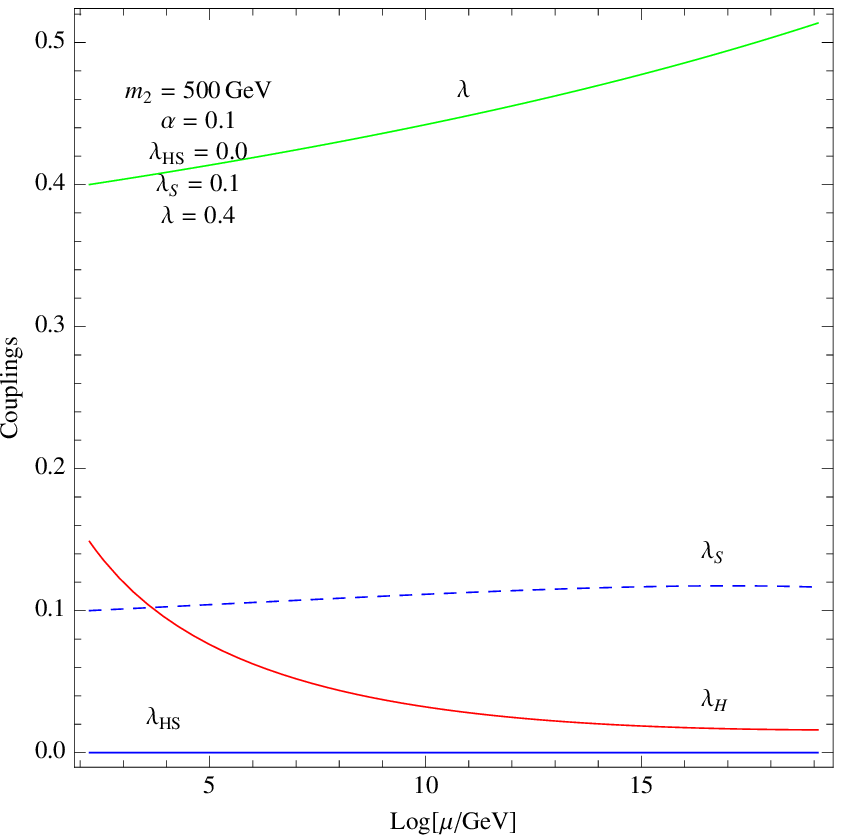}
\includegraphics[width=0.45\textwidth]{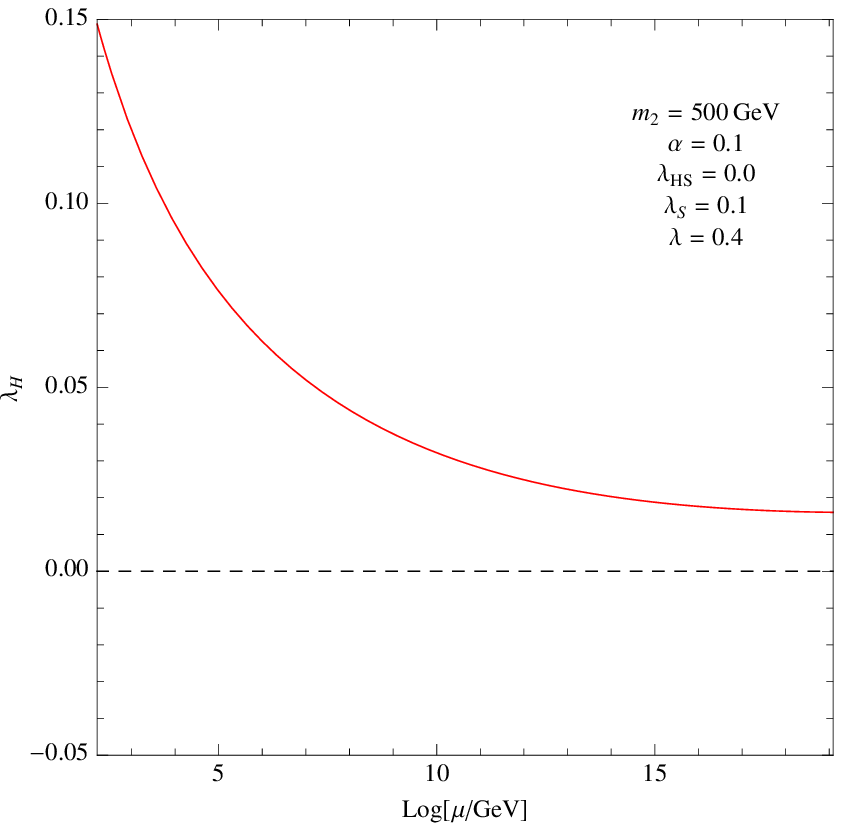}
\caption{RG-running of couplings as a function of renormalization scale for $m_1 = 125 \GeV$, $m_2=500 \GeV$ and $\alpha=0.1$, but $\lambda_{HS}=0$, i.e, mixing but no-loop correction. 
Red/blue/green/dashed-blue line corresponds to $\lambda_H$/$\lambda_{HS}$/$\lambda$/$\lambda_S$.}
\label{fig:runnings-mix}
\end{center}
\end{figure}

\begin{figure}[htbp]
\begin{center}
\includegraphics[width=0.45\textwidth]{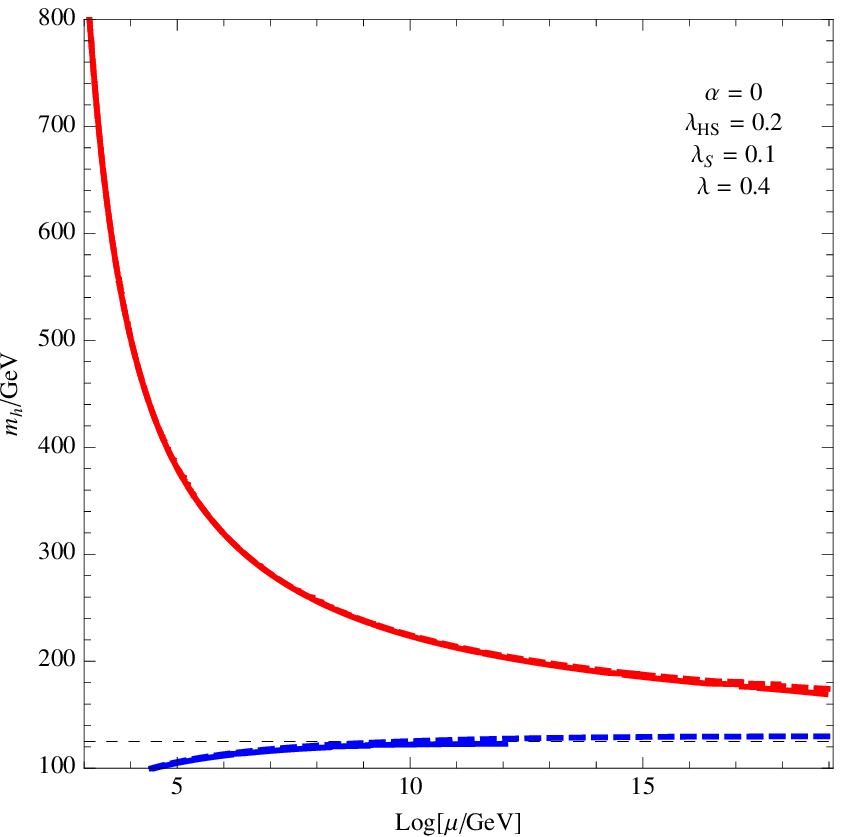}
\includegraphics[width=0.45\textwidth]{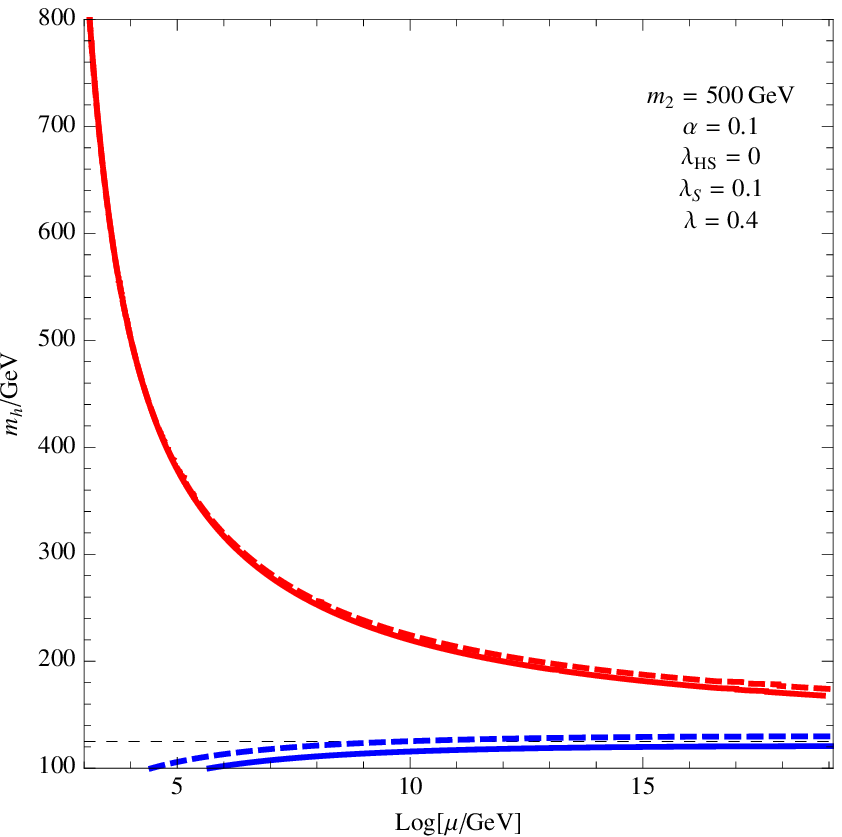}
\caption{The mass bound of SM-like Higgs ($m_1$) as a function of energy scale for $(\alpha,\lambda_{HS})=(0,0.2)$(left),$(0.1,0)$(right) with $\lambda_S = 0.1$ and $\lambda = 0.4$.
The red/blue line corresponds to triviality/vacuum-stability bound in SM(dashed) and our model(solid).
The dashed black line corresponds to $m_1 = 125 \GeV$.
}
\label{fig:mh-vs-mu}
\end{center}
\end{figure}

As an example parameter set for stable vacuum,  in Fig.~\ref{fig:runnings-nomix} and~\ref{fig:runnings-mix}, the RG-running of dimensionless couplings are shown for loop-effect only(left) and mixing only(right), respectively.
Comparing both figures, we observe that, because of increasing magnitude of $\lambda_{HS}$, the loop effect raises up the running of $\lambda_H$ at high renormalization scale.
Contrary to this, tree-level mixing effect is nearly like a simple shifting up of $\lambda_H$ that can be also seen clearly in Fig.~\ref{fig:mh-vs-mu}


The triviality and vacuum stability bound of Higgs mass as a function of the renormalization scale is shown in Fig.~\ref{fig:mh-vs-mu}. 
Dashed lines in the figure corresponds to standard model bounds.
Solid lines are bounds in our model.
Note that in the left panel where only loop effect is included the solid-blue instability line is terminated at an intermediate scale of $\mu$.
It is because the quartic self-interaction of Higgs decreases at low scale but increases at high scale, as can be seen in the right panel of Fig.~\ref{fig:runnings-nomix}
\footnote{
As can be seen from Fig.~\ref{fig:runnings-nomix}, if $\lambda_{HS}$ is small, $\lambda_H(\mu)$ can cross-down zero-point at a scale, but cross-up at a higher scale.
Instability, however, may not exist if SM-vacuum is the global minimum.
Even if SM-vacuum is metastable, it is okay as long as the tunneling time is longer than the age of our universe.
Cosmological danger of falling down to wrong vacuum may be also avoidable as long as Higgs field were in thermal touch with high temperature radiation background after inflation.
}.
As shown in the figure, both of loop effect of $\lambda_{HS}$ and tree-level mixing can remove the vacuum instability easily in our model though they work differently.

\subsection{Brief Summary}
In brief summary, the numerical analysis shows that the vacuum stability of 
Higgs potential and perturbativity of couplings constrains new dimensionless 
couplings of our model as follows.
\beq
0 \leq |\lambda| \lesssim 0.6 \ , \ -0.2 \lesssim \lambda_{HS} \lesssim 0.4 \ , \ 
0 \leq \lambda_S \lesssim 0.2 .
\label{coupling_summary}
\eeq
On the other hand, we cannot extract any useful bounds on the 2nd Higgs mass 
$m_2$,  since it depends on an unknown quantity, the singlet VEV $v_S$. 
\section{Conclusion}\label{sec:conclu}
In this paper, we have made a careful and comprehensive study of vacuum structure 
and vacuum stability in the singlet fermion dark matter with a real singlet scalar 
messenger. 
We found that the vacuum structure of this model has a very rich structure. 
In the parameter space we explore here, the EW, I, III and SYM phases (definitions are given in the text) appear, and any of them can become the global minimum.
It is remarkable that the region where the EW vacuum is the global minimum 
is overly limited, eliminating the large parameter space.
The tree-level analysis shows that the global vacuum condition yields
the constraint on the mass of the second Higgs boson, i.e., 
$\sqrt{\lambda_S/2}|v_S|\ltsim m_2\ltsim \sqrt{2\lambda_S}|v_S|$ 
for $v_S\gg v_H$ and $\alpha\ltsim$ 0.2 rad.

At the one-loop level, on top of the SM particles and singlet Higgs, 
the DM also participates in the effective potential.
We found that effects of the DM on the vacuum structure can be significant depending 
on the magnitude and sign of $\lambda$.
However, it tends to be small once we take $m_\psi=m_2/2$ which is favored by the 
DM relic abundance requirement.
In such a case, above tree-level bound on $m_2$ is still valid at the one-loop level 
as verified numerically. 
Our findings also applies to the SM extension with a singlet scalar boson without CDM.

We also studied vacuum instability caused by the RG-running effect of the top quark 
at high-energy scales.
Unlike the SM, $\lambda_H$ in the current model can be positive all the way to the Planck scale.
The reason is two-fold: due to the tree and loop effects coming from the singlet Higgs sector.

The former is related to the mixing between the doublet and singlet Higgs bosons at the tree level.
Because of this, the Higgs quartic coupling and mass is no longer one-to-one correspondence.
We can take larger $\lambda_H$ with $m_1=125$ GeV by increasing $\alpha$ and/or $m_2$
as shown in Eq.~(\ref{lambdaH-tree}).
This increment in the initial value of $\lambda_H$ can avoid zero-crossing up to the Planck scale.

The latter is nothing but the effect of $\lambda_{HS}$ on the RG-running of $\lambda_H$. 
Such an effect can be sizable enough to compensate the negative contribution coming from the top
quark at high-energy scale, preventing the Higgs potential from generating 
an unbounded-from-below direction or a potentially new global minimum.
The perturbativity of the quartic couplings up to the Planck scale 
are also investigated. The results are summarized in Eq.~(\ref{coupling_summary}). 

Before closing, we make a comment on the bound of $m_2$.
Although we obtain the strong bound on $m_2$ through the global vacuum condition,
$m_2$ is still not predictable since $v_S$ and $\lambda_S$ are totally unknown.
Therefore, we may need additional information to constrain those parameters somehow.
For example, since strong first-order electroweak phase transition as needed for successful 
electroweak baryogenesis is closely related to specific vacuum structure 
and so the specific $v_S$ and $\lambda_S$, we may get
more useful bound on $m_2$. We leave this possibility to a future study.

\section{Note Added}

\begin{figure}
\centering
\includegraphics[width=0.7\textwidth]{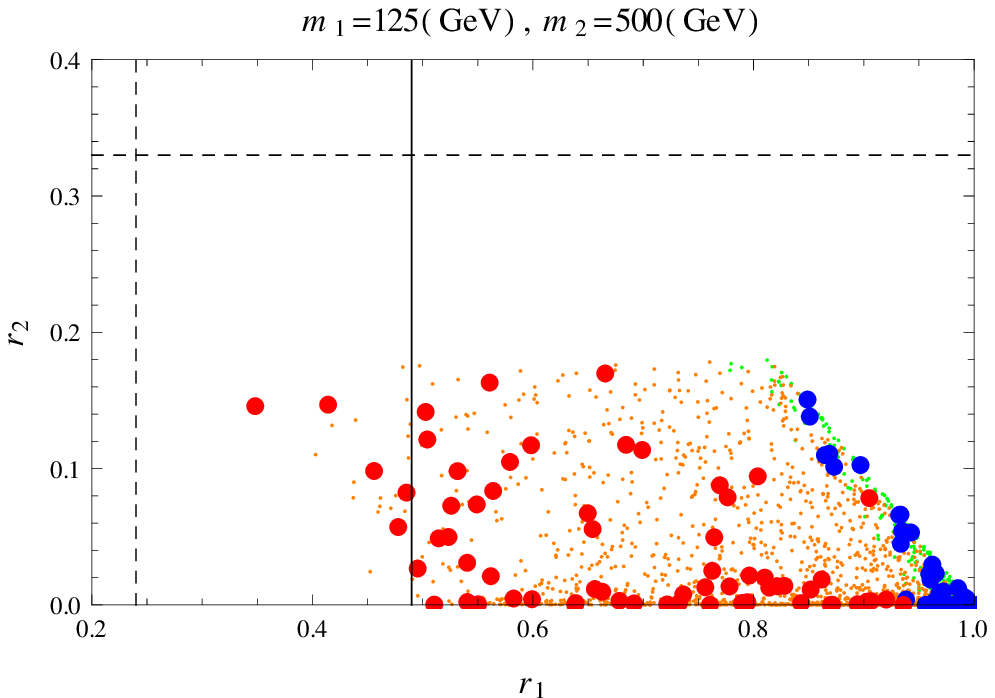}
\includegraphics[width=0.7\textwidth]{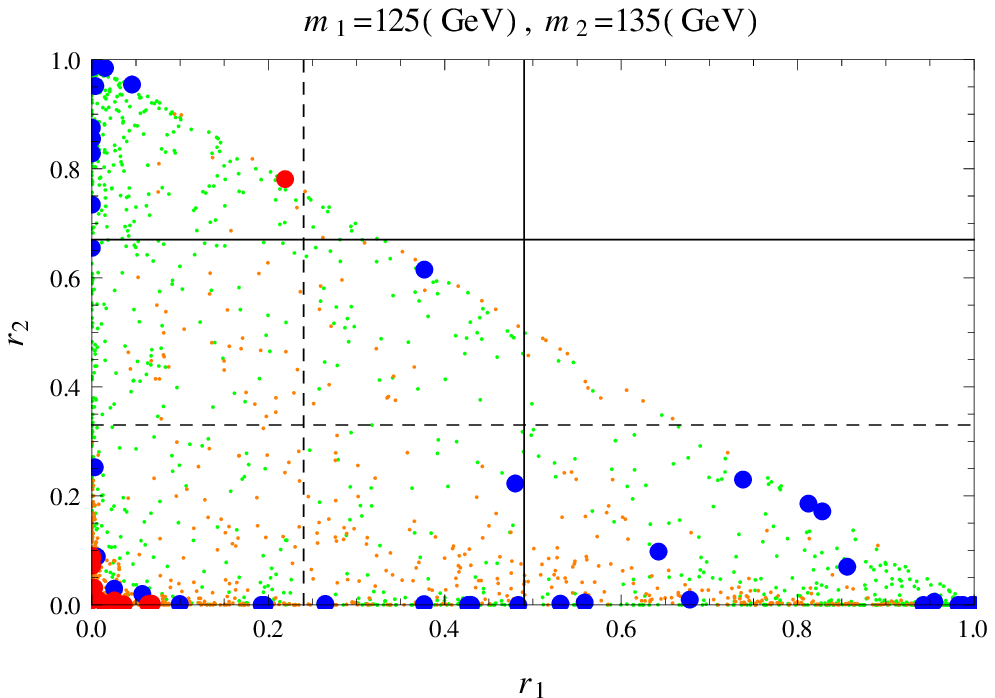}
\includegraphics[width=0.7\textwidth]{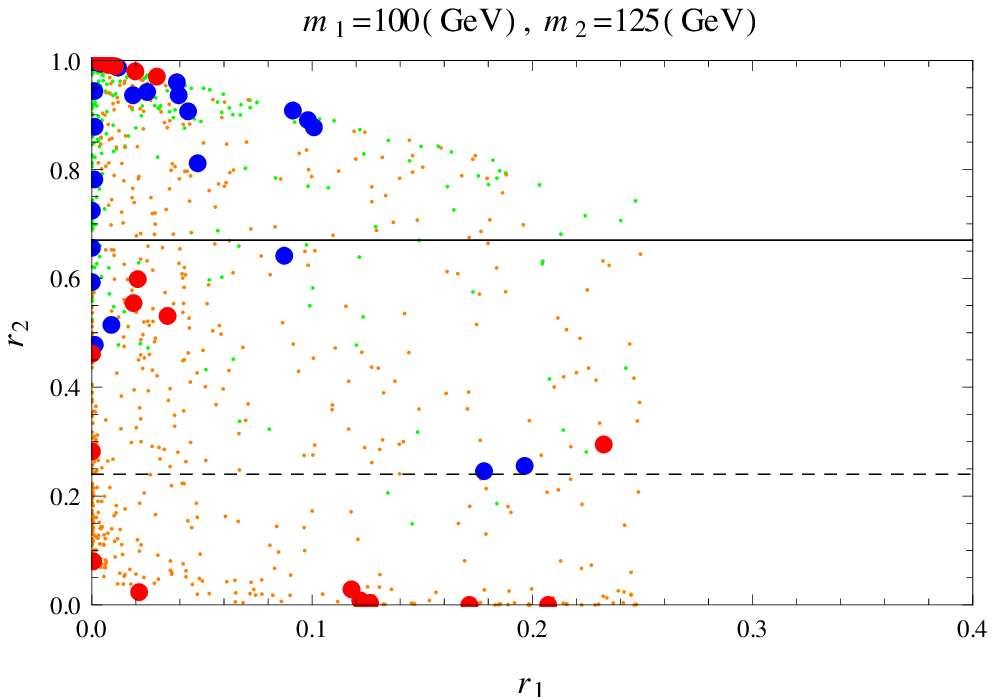}
\caption{Scatter plot in $(r_1,r_2)$ plane for the scenario S1, S2 and S3 (from above). 
The points represent 4 different cases:
$(\Om_{\rm CDM} h^2)^{3 \si}$, $\si_p^>$ (big red),
$(\Om_{\rm CDM} h^2)^{3 \si}$, $\si_p^<$ (big blue),
$(\Om_{\rm CDM} h^2)^{<}$, $\si_p^>$ (small orange),
and $(\Om_{\rm CDM} h^2)^{<}$, $\si_p^<$ (small green). 
(See the text for more detail).
}
\label{fig:r1r2}
\end{figure}

While we are finalizing this work, CMS and ATLAS Collaborations reported 
a new boson of mass around 125 GeV, which might be consistent with 
the SM Higgs boson. 
The invisible branching ratio of the observed particle seems to be small.
In Fig.~\ref{fig:r1r2}, we show the reduction factor $r_i$ for one of the Higgs 
boson mass is equal to 125 GeV.  
(Note that we took $m_h = 120$ GeV in Ref.~\cite{Baek:2011aa}.)

\begin{figure}
\centering
\includegraphics[width=0.8\textwidth]{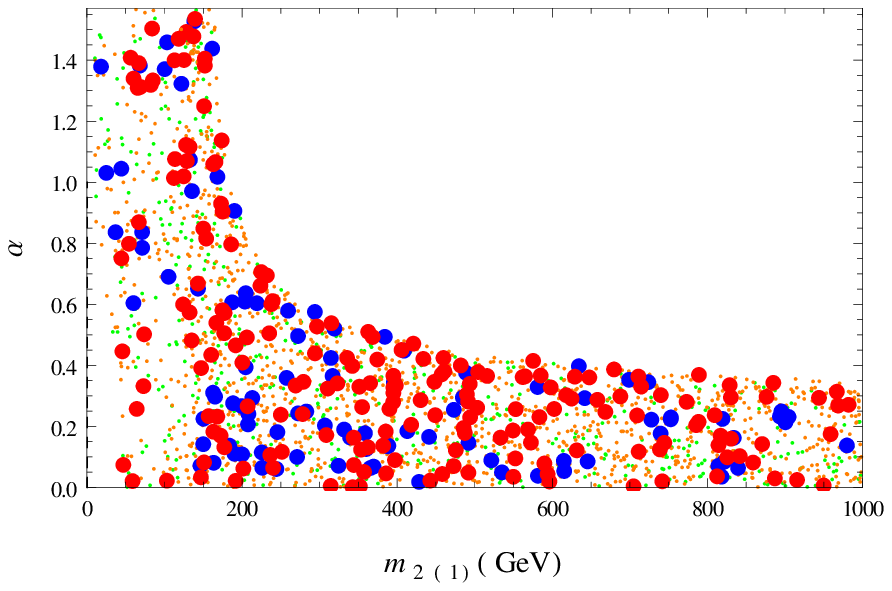}
\caption{Scattered plot in $(m_{2(1)},\alpha)$ plane for 
$m_2 > m_1 =125~\GeV (m_1 < m_2 =125~\GeV)$. 
The color scheme is the same with Fig.~\ref{fig:r1r2}.
}
\label{fig:m2_alpha}
\end{figure}

Implications of this new results on the model studied in this work 
are the following:
\begin{itemize}
\item Since the observed properties of a new boson is close to those of the SM 
Higgs boson, its singlet component should be small, namely the mixing angle 
$\alpha$ should be small in our model.
\item If $m_{1} = 125$ GeV, the singlet fermion DM mass may have to be greater
than $m_{1} /2 \sim 63$ GeV so that $H_1 \rightarrow \psi \overline{\psi}$ 
is kinematically forbidden. The current Higgs search results can not be directly 
applied to $H_2$,  since $H_2$ would be mostly a singlet and difficult to be produced
at colliders. Also it may decay into a pair of DM's with a substantial branching ratio.
Note that $r_2 < 0.3$ from Fig. 10 of Ref.~\cite{Baek:2011aa}.  
\item If $m_{2} = 125$ GeV, the lighter Higgs boson $H_1$ can be light and may have 
escaped the detection if it is mostly SM singlet. This is still consistent with all the 
data available as of now. 
\item In order to test the idea of Higgs portal dark matter with a singlet fermion
DM, it is important to search for two Higgs-like scalar bosons, one of which is to
be identified with a new particle with mass around 125 GeV. The other scalar 
could be heavy and have escaped the Higgs search at the LHC and at the Tevatron,
if it has substantial invisible branching ratio into a pair of DM's. 
\item $r_i$ defined in Eq.~(3.4) is always less than one in our model. 
Therefore, our model would be excluded, if $r_i$ turns out to be larger than 
one ($r_i  > 1$) in any of the decay channels in the future. 
On the other hand, if $r_q < 1$ is observed in all observed channels in the 
future,  our model could be a good candidate for the reason behind it.
We have to wait for more precise determinations of $r_i$ for all possible 
measurable decay channels of Higgs-like boson with 125 GeV mass.
\item In our model, there are two Higgs-like scalar bosons, one of which could be
mostly singlet scalar and hard to discover at colliders, since $r_i < 1$. 
In Fig.~\ref{fig:m2_alpha}, we show a scattered plot in 
the $(m_{2(1)},\alpha)$-plane for $m_2 > m_1 =125~\GeV (m_1 < m_2 =125~\GeV)$.
Note that there is no strong constraint on the allowed mass range for the 2nd Higgs
from the EWPT, collider or DM phenomenology. And the current SM Higgs search bounds
do not apply directly, since the signal strength is reduced in our model ($r_i < 1$) 
compared with the SM Higgs boson.  Note that $\alpha \lesssim 0.4 \times \pi/2$ for 
all values of $m_2 \gtrsim 220~\GeV$,  for which case $H_2$ is always singlet-like. 
We find that $r_2 \lesssim 0.4$ for  $m_2 \gtrsim 220~\GeV$, making it difficult to 
search for the heavier (singlet-like)  scalar particle.  
Therefore the current Higgs search should be  continued for wider ranges of Higgs 
mass considering a possibility of  $r_i <1$,  
in order to look whether another Higgs-like scalar boson (mostly singlet-like) exists or not. 
\end{itemize}

While we are finalizing this paper, there appeared a paper which considers 
Higgs phenomenology of a similar model, a singlet extension of the SM with DM 
in a hidden sector~\cite{Batell:2012mj}. 

\section*{Acknowledgements}

This work is supported in part by NRF Research Grant 
2012R1A2A1A01006053 (PK and SB), and by SRC program of NRF 
Grant No. 20120001176 funded by MEST through 
Korea Nneutrino Research Center at Seoul National University (PK).


\appendix
\section{The top-quark running mass}\label{app:top_quark}
The top-quark pole-mass is related to the $\overline{\rm MS}$ mass $m_t(\mu)$ to NNLO as
\beq
M_t = m_t(\mu) \l[ 1 + \l(\frac{\alpha_s^{(n_f=5)}(\mu)}{\pi} \r) d_1 + \l(\frac{\alpha_s^{(n_f=5)}(\mu)}{\pi} \r)^2 d_2 \r]
\eeq
where $\alpha_s^{(n_f=5)}$ is the strong coupling with five active flavors, $d_i$s are given as \cite{Langenfeld:2009wd}
\bea
d_1 &=& \frac{4}{3} + \ln \frac{\mu^2}{m_t(\mu)^2}
\\
d_2 &\simeq& \frac{307}{32} + 2 \zeta_2 + \frac{3}{2} \zeta_2 \ln 2 - \frac{1}{6} \zeta_3 + \frac{509}{72} \ln \frac{\mu^2}{m_t(\mu)^2} + \frac{47}{24} \ln^2 \frac{\mu^2}{m_t(\mu)^2}
\nonumber \\
&& - \l( \frac{71}{144} + \frac{1}{3} \zeta_2 + \frac{13}{36}  \ln \frac{\mu^2}{m_t(\mu)^2} + \frac{1}{12}  \ln^2 \frac{\mu^2}{m_t(\mu)^2} \r) n_f
\eea
with $\zeta_2 \approx 1.645$ and $\zeta_3 \approx 1.202$ being zeta-constants.
For matching, we have to set $\mu = m_t$.
At this scale the strong coupling can be found from the following relation for the running coupling
\bea
\alpha_s^{(n_f=5)}(m_t) 
&=& 
\alpha_s^{(n_f=5)}(\mu) \l[ 1 + 4 \pi \alpha_s^{(n_f=5)}(\mu) \beta_0 \ln \frac{\mu^2}{m_t(\mu)^2} \r.
\nonumber \\
&& \l. + \l( 4 \pi \alpha_s^{(n_f=5)}(\mu) \r)^2 \l( \beta_1 \ln \frac{\mu^2}{m_t(\mu)^2}  + \beta_0^2 \ln^2 \frac{\mu^2}{m_t(\mu)^2} \r) \r]
\eea
where
\bea
\beta_0 &=& \frac{1}{16 \pi^2} \l( 11 - \frac{2}{3} n_f \r)
\\
\beta_1 &=& \frac{1}{\l( 16 \pi^2 \r)^2} \l( 102 - \frac{38}{3} n_f \r).
\eea
\section{Higgs boson masses at the one-loop level}\label{app:1LmH}
\label{app:1LHiggsMass}
\allowdisplaybreaks{
Here, we write the useful formulae for the Higgs boson masses at the one-loop level.
Let us begin by listing the field-dependent masses of all particles.
The field-dependent masses of the Higgs bosons are given by the eigenvalues of the following
2-by-2 mass matrix
\begin{align}
\bar{M}_{\rm Higgs}^2(\varphi_H,\varphi_S) &= 
\left(
	\begin{array}{cc}
	-\mu_H^2+3\lambda_H\varphi_H^2+\mu_{HS}\varphi_S+\frac{\lambda_{HS}}{2}\varphi_S^2
	& \mu_{HS}\varphi_H+\lambda_{HS}\varphi_H\varphi_S \\
	\mu_{HS}\varphi_H+\lambda_{HS}\varphi_H\varphi_S 
	& m_S^2+2\mu'_S\varphi_S+3\lambda_S\varphi_S^2+\frac{\lambda_{HS}}{2}\varphi_H^2
	\end{array}
\right) \nonumber\\
&=
\left(
	\begin{array}{cc}
	\bar{m}_{hh}^2 & \bar{m}_{hs}^2 \\
	\bar{m}_{hs}^2 & \bar{m}_{ss}^2
	\end{array}
\right).
\end{align}
The field-dependent masses of the NG bosons, gauge bosons, top/bottom and singlet fermionic DM
are respectively given by
\begin{align}
\bar{m}_{G^0}^2(\varphi_H, \varphi_S)&=\bar{m}_{G^\pm}^2(\varphi_H, \varphi_S)
 = -\mu_H^2+\lambda_H\varphi_H^2+\mu_{HS}\varphi_S+\frac{\lambda_{HS}}{2}\varphi_S^2, \\
\bar{m}_W^2(\varphi_H) &=\frac{g_2^2}{4}\varphi_H^2, \quad
\bar{m}_Z^2(\varphi_H) =\frac{g_2^2+g_1^2}{4}\varphi_H^2, \\
\bar{m}_t^2(\varphi_H) &= \frac{\lambda_t^2}{2}\varphi_H^2,\quad
\bar{m}_b^2(\varphi_H) = \frac{\lambda_b^2}{2}\varphi_H^2, \\
\bar{m}_\psi^2(\varphi_S) &= m_{\psi_0}^2+2\lambda m_{\psi_0}\varphi_S+\lambda^2\varphi_S^2.
\end{align}
The one-loop tadpole conditions are cast into the form
\begin{align}
\frac{1}{v_H}\left\langle\frac{\partial V_{\rm eff}}{\partial \varphi_H} \right\rangle &=
	-\mu_H^2+\lambda_Hv_H^2+\mu_{HS}v_S+\frac{\lambda_{HS}}{2}v_S^2 \nonumber\\
&\quad +\frac{1}{16\pi^2}
\bigg[
	\frac{6\lambda_H+\lambda_{HS}}{4}f_+(m_1^2,m_2^2)\nonumber\\
&\hspace{2cm}	
	-\frac{1}{4\Delta m_H^2}
	\Big\{
		(6\lambda_H-\lambda_{HS})(m_{hh}^2-m_{ss}^2)+4(\mu_{HS}+\lambda_{HS}v_S)^2
	\Big\}f_-(m_1^2,m_2^2) \nonumber\\
&\hspace{2cm}	
	+\lambda_H
	\left\{
		m_{G^0}^2\left(\ln\frac{m_{G^0}^2}{\mu^2}-1\right)
		+2m_{G^\pm}^2\left(\ln\frac{m_{G^\pm}^2}{\mu^2}-1\right)	
	\right\} \nonumber\\
&\hspace{2cm}	
	+\frac{3}{v_H^2}\left\{ 
		2m_W^4\left(\ln\frac{m_W^2}{\mu^2}-\frac{1}{3}\right)
		+m_Z^4\left(\ln\frac{m_Z^2}{\mu^2}-\frac{1}{3}\right)
	\right\} \nonumber\\
&\hspace{2cm}
	-\frac{12}{v_H^2}
	\left\{
		m_t^4\left(\ln\frac{m_t^2}{\mu^2}-1\right)
		+m_b^4\left(\ln\frac{m_b^2}{\mu^2}-1\right)
	\right\}
\bigg]
= 0, \\
\frac{1}{v_S}\left\langle\frac{\partial V_{\rm eff}}{\partial \varphi_S} \right\rangle &= 
	\frac{\mu_S^3}{v_S}+m_S^2+\mu'_Sv_S+\lambda_Sv_S^2
	+\frac{\mu_{HS}}{2}\frac{v_H^2}{v_S}+\frac{\lambda_{HS}}{2}v_H^2 \nonumber \\
&\quad +\frac{1}{16\pi^2}
\bigg[
	\frac{1}{4}\left(\frac{\mu_{HS}+2\mu'_S}{v_S}+\lambda_{HS}+6\lambda_S\right)f_+(m_1^2,m_2^2)\nonumber\\
&\hspace{2cm}	
	-\frac{1}{4\Delta m_H^2}
	\bigg\{
		\left(\frac{\mu_{HS}-2\mu'_S}{v_S}+\lambda_{HS}-6\lambda_S\right)
		(m_{hh}^2-m_{ss}^2) \nonumber\\
&\hspace{6cm}
	+4\lambda_{HS}v_H^2\left(\frac{\mu_{HS}}{v_S}+\lambda_{HS}\right)
	\bigg\}f_-(m_1^2,m_2^2) \nonumber\\
&\hspace{2cm}	
	+\frac{1}{2}\left(\frac{\mu_{HS}}{v_S}+\lambda_{HS}\right)
	\left\{
		m_{G^0}^2\left(\ln\frac{m_{G^0}^2}{\mu^2}-1\right)
		+2m_{G^\pm}^2\left(\ln\frac{m_{G^\pm}^2}{\mu^2}-1\right)	
	\right\} \nonumber\\	
&\hspace{2cm}
	-4\lambda\left(\frac{m_{\psi_0}}{v_S}+\lambda\right)m_{\psi}^2
	\left(\ln\frac{m_{\psi}^2}{\mu^2}-1\right)
\bigg]
= 0,\label{1L-tad_bvS}
\end{align}
where
\begin{align}
 m_i^2&=\langle \bar{m}_i^2\rangle = \bar{m}_i^2(v_H,v_S), \\
f_\pm(m_1^2, m_2^2) &= 
	m_1^2\left(\ln\frac{m_1^2}{\mu^2}-1\right)
	\pm m_2^2\left(\ln\frac{m_2^2}{\mu^2}-1\right), \\
\Delta m_H^2 &= m_2^2-m_1^2= \sqrt{(m_{hh}^2-m_{ss}^2)^2+4m_{hs}^4}.
\end{align}
After imposing the one-loop tadpole conditions, the mass matrix elements of the 
Higgs bosons are
\begin{align}
(M_{\rm Higgs}^2)_{11}
&= 2\lambda_Hv_H^2 +\frac{v_H^2}{32\pi^2}
\bigg[
	-A_1f_-(m_1^2,m_2^2)+A_2\ln\frac{m_1^2m_2^2}{\bar{\mu}^4}
	-A_3\ln\frac{m_1^2}{m_2^2}\nonumber\\
&\hspace{3.5cm} +4\lambda_H^2
	\left(\ln\frac{m_{G^0}^2}{\mu^2}+2\ln\frac{m_{G^\pm}^2}{\mu^2} \right) \nonumber\\
&\hspace{3.5cm}
	+\frac{3}{4}
	\left\{
		2g_2^4\left(\ln\frac{m_W^2}{\mu^2}+\frac{2}{3}\right)
		+(g_2^2+g_1^2)^2\left(\ln\frac{m_Z^2}{\mu^2}+\frac{2}{3}\right)
	\right\} \nonumber\\
&\hspace{3.5cm}
	-12\left(\lambda_t^4\ln\frac{m_t^2}{\mu^2}+\lambda_b^4\ln\frac{m_b^2}{\mu^2} \right)
\bigg], \\
(M_{\rm Higgs}^2)_{22}
&= -\frac{\mu_S^3}{v_S}+\mu'_Sv_S+2\lambda_Sv_S^2-\frac{\mu_{HS}}{2}\frac{v_H^2}{v_S}\nonumber\\
&\quad +\frac{v_S^2}{32\pi^2}
\Bigg[
	B_0f_+(m_1^2,m_2^2)-B_1f_-(m_1^2,m_2^2)
	+B_2\ln\frac{m_1^2m_2^2}{\bar{\mu}^4}
	-B_3\ln\frac{m_1^2}{m_2^2}\nonumber\\
&\hspace{2cm}
	-\frac{\mu_{HS}}{v_S^3}m_{G^0}^2\left(\ln\frac{m_{G^0}^2}{\mu^2}-1\right)
	+\left(\frac{\mu_{HS}}{v_S}+\lambda_{HS}\right)^2\ln\frac{m_{G^0}^2}{\mu^2} \nonumber\\
&\hspace{2cm}
	+2\left\{
	-\frac{\mu_{HS}}{v_S^3}m_{G^\pm}^2\left(\ln\frac{m_{G^\pm}^2}{\mu^2}-1\right)
	+\left(\frac{\mu_{HS}}{v_S}+\lambda_{HS}\right)^2\ln\frac{m_{G^\pm}^2}{\mu^2}
	\right\} \nonumber\\
&\hspace{2cm}-4
	\left\{
		-\frac{2\lambda m_{\psi_0}}{v_S^3}m_{\psi}^2
		\left(\ln\frac{m_{\psi}^2}{\mu^2}-1\right)
		+4\lambda^2\left(\frac{m_{\psi_0}}{v_S}
		+\lambda\right)^2\ln\frac{m_{\psi}^2}{\mu^2}
	\right\}	
\Bigg], \\
(M_{\rm Higgs}^2)_{12}&=(M_{\rm Higgs}^2)_{21} \nonumber\\
&=\mu_{HS}v_H+\lambda_{HS}v_Hv_S \nonumber\\
&\quad
+\frac{v_Hv_S}{32\pi^2}
\bigg[
	-C_1f_-(m_1^2,m_2^2)
	+C_2\ln\frac{m_1^2m_2^2}{\bar{\mu}^4}
	-C_3\ln\frac{m_1^2}{m_2^2}\nonumber\\
&\hspace{2cm}
	+2\lambda_H\left(\frac{\mu_{HS}}{v_S}+\lambda_{HS}\right)
	\left(\ln\frac{m_{G^0}^2}{\mu^2}+2\ln\frac{m_{G^\pm}^2}{\mu^2}\right)
\bigg], 
\end{align}
where
\begin{align}
A_1 &= \frac{1}{2\Delta m_H^2}
\left[
	(6\lambda_H-\lambda_{HS})^2-\frac{1}{(\Delta m_H^2)^2}
	\Big\{
		(6\lambda_H-\lambda_{HS})(m_{hh}^2-m_{ss}^2)+4(\mu_{HS}+\lambda_{HS}v_S)^2
	\Big\}^2
\right], \\
A_2 &= \frac{1}{4}(6\lambda_H+\lambda_{HS})^2
	+\frac{1}{4(\Delta m_H^2)^2}
	\Big\{
		(6\lambda_H-\lambda_{HS})(m_{hh}^2-m_{ss}^2)
		+4(\mu_{HS}+\lambda_{HS}v_S)^2
	\Big\}^2, \\
A_3 &= \frac{1}{2\Delta m_H^2}(6\lambda_H+\lambda_{HS})
	\Big\{
		(6\lambda_H-\lambda_{HS})(m_{hh}^2-m_{ss}^2)
		+4(\mu_{HS}+\lambda_{HS}v_S)^2
	\Big\}, \\
B_0 &= -\frac{\mu_{HS}+2\mu'_S}{2v_S^3}, \\
B_1 &= \frac{1}{2\Delta m_H^2}
\left[
	\left(\frac{\mu_{HS}-2\mu'_{S}}{v_S}+\lambda_{HS}-6\lambda_S\right)^2
	-\frac{1}{v_S^2}
	\left\{
		\frac{\mu_{HS}-2\mu'_{S}}{v_S}(m_{hh}^2-m_{ss}^2)
		+4\lambda_{HS}\mu_{HS}\frac{v^2}{v_S}
	\right\}
\right] \nonumber\\
&\quad -\frac{1}{2(\Delta m_H^2)^3}
\left[
	\left(\frac{\mu_{HS}-2\mu'_{S}}{v_S}+\lambda_{HS}-6\lambda_S\right)(m_{hh}^2-m_{ss}^2)
	+4\lambda_{HS}v_H^2\left(\frac{\mu_{HS}}{v_S}+\lambda_{HS}\right)
\right]^2, \\
B_2 &= \frac{1}{4}\left(\frac{\mu_{HS}+2\mu'_{S}}{v_S}+\lambda_{HS}+6\lambda_S\right)^2
	\nonumber\\
&\quad +\frac{1}{4(\Delta m_H^2)^2}
\left[
	\left(
		\frac{\mu_{HS}-2\mu'_{S}}{v_S}+\lambda_{HS}-6\lambda_S
	\right)(m_{hh}^2-m_{ss}^2)
	+4\lambda_{HS}v_H^2\left(\frac{\mu_{HS}}{v_S}+\lambda_{HS}\right)
\right]^2, \\
B_3 &= \frac{1}{2\Delta m_H^2}
	\left(\frac{\mu_{HS}+2\mu'_{S}}{v_S}+\lambda_{HS}+6\lambda_S\right) \nonumber\\
&\hspace{2cm}\times
\left[
	\left(
		\frac{\mu_{HS}-2\mu'_{S}}{v_S}+\lambda_{HS}-6\lambda_S
	\right)(m_{hh}^2-m_{ss}^2)
	+4\lambda_{HS}v_H^2\left(\frac{\mu_{HS}}{v_S}+\lambda_{HS}\right)
\right], \\
C_1 &= \frac{1}{2\Delta m_H^2}
\left[
	(6\lambda_H-\lambda_{HS})
	\left(\frac{\mu_{HS}-2\mu'_S}{v_S}+\lambda_{HS}-6\lambda_S \right)
	+8\lambda_{HS}\left(\frac{\mu_{HS}}{v_S}+\lambda_{HS}\right)
\right] \nonumber\\
&\quad -\frac{1}{2(\Delta m_H^2)^3}
\Big[
	(6\lambda_H-\lambda_{HS})(m_{hh}^2-m_{ss}^2)+4(\mu_{HS}+\lambda_{HS}v_S)^2
\Big] \nonumber\\
&\hspace{3cm} \times
\left[
	\left(\frac{\mu_{HS}-2\mu'_S}{v_S}+\lambda_{HS}-6\lambda_S \right)(m_{hh}^2-m_{ss}^2)
	+4\lambda_{HS}v_H^2\left(\frac{\mu_{HS}}{v_S}+\lambda_{HS}\right)
\right], \\
C_2 &= \frac{1}{4}(6\lambda_H+\lambda_{HS})
	\left(\frac{\mu_{HS}+2\mu'_S}{v_S}+\lambda_{HS}+6\lambda_S \right) \nonumber\\
&\quad +\frac{1}{4(\Delta m_H^2)^2}
	\Big[
		(6\lambda_H-\lambda_{HS})(m_{hh}^2-m_{ss}^2)
		+4(\mu_{HS}+\lambda_{HS}v_S)^2
	\Big] \nonumber\\
&\hspace{3cm} \times
	\left[
		\left(\frac{\mu_{HS}-2\mu'_S}{v_S}+\lambda_{HS}-6\lambda_S \right)(m_{hh}^2-m_{ss}^2)
		+4\lambda_{HS}v_H^2\left(\frac{\mu_{HS}}{v_S}+\lambda_{HS}\right)
	\right], \\
C_3 &= \frac{1}{4\Delta m_H^2}	
\bigg[
	(6\lambda_H+\lambda_{HS})
	\left\{
		\left(\frac{\mu_{HS}-2\mu'_S}{v_S}+\lambda_{HS}-6\lambda_S \right)(m_{hh}^2-m_{ss}^2)
		+4\lambda_{HS}v_H^2\left(\frac{\mu_{HS}}{v_S}+\lambda_{HS}\right)
	\right\} \nonumber\\
&\hspace{2cm}
	+\left(\frac{\mu_{HS}+2\mu'_S}{v_S}+\lambda_{HS}+6\lambda_S \right)
	\Big\{
		(6\lambda_H-\lambda_{HS})(m_{hh}^2-m_{ss}^2)+4(\mu_{HS}+\lambda_{HS}v_S)^2
	\Big\}	
\bigg].
\end{align}
}

\end{document}